\documentclass[aps,prd,epsfigure,twocolumn,superscriptaddress,longbibliography,nofootinbib]{revtex4-1}
\usepackage{dcolumn}    
\usepackage{bm} 
\usepackage{graphicx}
\usepackage{amsmath}    
\usepackage{latexsym}
\usepackage{amsfonts}   
\usepackage{amssymb}
\usepackage{array}      
\usepackage{epsfig}
\usepackage{txfonts}
\usepackage{color}
\usepackage[colorlinks=true,linkcolor=blue,urlcolor=blue,citecolor=blue,pdfusetitle]{hyperref}
\usepackage{hyperref}

\usepackage{titlesec}

\newcommand{\be}{\begin{equation}}
\newcommand{\ee}{\end{equation}}
\newcommand{\baln}{\begin{aligned}}
\newcommand{\ealn}{\end{aligned}}
\newcommand{\ben}{\begin{equation*}}
\newcommand{\een}{\end{equation*}}

\newcommand{\ket}[1]{\left\vert#1\right\rangle}

\newcommand{\blah}{blah\\blah\\blah\\blah\\blah.}

\newcommand{\del}{\partial}

\begin{document}

\title{Tests of Quantum Gravity-Induced Non-Locality:\\ Hamiltonian formulation of a non-local harmonic oscillator}

\author{A. Belenchia}
\affiliation{Centre for Theoretical Atomic, Molecular, and Optical Physics, School of Mathematics and Physics, Queens University, Belfast BT7 1NN, United Kingdom}
\author{D. Benincasa}
\affiliation{School of Theoretical Physics, Dublin Institute for Advanced Studies, 10 Burlington Road, Dublin 4, Ireland}
\author{F. Marin}
\affiliation{Dipartimento di Fisica e Astronomia, University of 
Florence, Via Sansone 1, I-50019 Sesto Fiorentino (FI), Italy.}
\affiliation{CNR-Istituto Nazionale di Ottica, Largo E. Fermi 6, 
I-50125 Firenze, Italy.}
\affiliation{INFN, Sez. di Firenze., Via Sansone 1, I-50019 Sesto 
Fiorentino (FI), Italy}
\affiliation{European Laboratory for Non-Linear Spectroscopy (LENS), 
Via Carrara 1, I-50019 Sesto Fiorentino (FI), Italy.}
\author{F. Marino}
\affiliation{CNR-Istituto Nazionale di Ottica, Largo E. Fermi 6, 
I-50125 Firenze, Italy.}
\affiliation{INFN, Sez. di Firenze., Via Sansone 1, I-50019 Sesto 
Fiorentino (FI), Italy}
\author{A. Ortolan}
\affiliation{INFN, Laboratori Nazionali di Legnaro, Viale dell'Universit\`a, 2, 35020 Legnaro, Padova, Italy}
\author{M. Paternostro}
\affiliation{Centre for Theoretical Atomic, Molecular, and Optical Physics, School of Mathematics and Physics, Queens University, Belfast BT7 1NN, United Kingdom}
\author{S. Liberati}
\affiliation{SISSA, Via Bonomea 265, I-34136 Trieste, Italy and INFN, Sez. di Trieste}

\begin{abstract}
Motivated by the development of on-going optomechanical experiments aimed at constraining non-local effects inspired by some quantum gravity scenarios, the Hamiltonian formulation of a non-local harmonic oscillator, and its coupling to a cavity field mode(s), is investigated. In particular, we consider the previously studied model of non-local oscillators obtained as the non-relativistic limit of a class of non-local Klein-Gordon operators, $f(\Box)$, with $f$ an analytical function. The results of previous works, in which the interaction was not included, are recovered and extended by way of standard perturbation theory. At the same time, the perturbed energy spectrum becomes available in this formulation, and we obtain the Langevin's equations characterizing the interacting system. 
\end{abstract}
\maketitle

\section{Introduction}\label{I}
The search for a quantum theory of gravity is plagued by lack of direct experimental tests and thus faces the challenge of testing low-energy predictions of theories otherwise empirically unverifiable. This task led to the blossoming in the last two decades of a new field of research, often referred to as Quantum Gravity Phenomenology~\cite{hossenfelder2010experimental,liberati2011quantum,AmelinoCamelia:2002vw}. Such a denomination collects under its umbrella different communities and different approaches to the same problem of testing the quantum features, if any, of gravity and spacetime.

In the search for possible remnant effects of Quantum Gravity (QG) at low-energy, a well-established body of research has focused on tests of Lorentz Invariance violations, leading to several strong constraints~\cite{Liberati:2013xla} (at least for its effective field theory formulation). Also, thanks to these results, growing attention has been drawn to alternative models which respect Lorentz symmetry in their low energy, large scale limit~\cite{AmelinoCamelia:2002vw,PhysRevLett.116.161303,Belenchia:2016sym,sorkin2009does,barnaby2008dynamics,Biswas:2014yia,Gambini:2014kba,Saravani:2015rva,PhysRevD.71.084012}. Some non-local field theories~\cite{Belenchia:2014fda,barnaby2008dynamics,Hossenfelder:2007re,Modesto:2017sdr,Barci:1995ad,Arzano:2018gii,Tomboulis:2015gfa}, i.e.~theories characterized by kinetic operators with infinitely many derivatives, fall in this category. 
The underlying idea here is that, preserving fundamental Lorentz invariance (LI) while accounting for the emergence of the continuum spacetime  from more fundamental constituents, leads to low energy effective theories with non-local dynamics. Relevant examples in the QG's literature exist in causal set theory~\cite{sorkin2009does,Belenchia:2014fda}; string theory and string field theory~\cite{Taylor:2003gn,barnaby2008dynamics}; and noncommutative geometry~\cite{Szabo:2001kg}. More in general, imposing LI and requiring the dynamics to be stable effectively single out non-local modifications of the local dynamics in order to avoid Ostrogardsky-like instabilities~\cite{ostrogradsky1850memoires}. Furthermore,  it is crucial to note that the scale which characterizes the non-locality does not necessarily correspond to the Planck scale, as it is indeed the case in the QG examples mentioned. This fact is of particular relevance for phenomenological studies aiming to cast constrains on this  mesoscopic scale, see, e.g., results obtained exploiting LHC data~\cite{Biswas:2014yia}.

In~\cite{PhysRevLett.116.161303}, the non-relativistic limit of a non-local Klein-Gordon field was derived, and the corresponding non-local Schr\"odinger equation was solved perturbatively. The main aim was to compare the effects of the non-local dynamics to realistic optomechanical set-ups, in order to bound the non-locality scale by means of table-top experiments in the near future. In particular, it was shown that table-top experiments have the potential to outperform high-energy constraints of non-local field theories currently based on LHC's data~\cite{Biswas:2014yia}.

Albeit promising, the analysis of~\cite{PhysRevLett.116.161303,PhysRevD.95.026012} was based on solving the Schr\"odinger equation perturbatively by choosing \textit{ad hoc} initial conditions, an ansatz for the form of the solutions, and it did not contain the interaction with external degrees of freedom, limiting \textit{de facto} the type of experimental protocols that can be used. 

A step forward with respect to these limitations is taken in the present work, in which we shall derive the Hamiltonian formulation of the non-local harmonic oscillator model investigated in~\cite{PhysRevLett.116.161303,PhysRevD.95.026012}. Obtaining the Hamiltonian will allow us to perform a new perturbative study of the solutions of the non-local oscillator by way of standard time-dependent and time-independent perturbation theory. In particular, we shall be able to go beyond what was previously done, by choosing more realistic preparation states, which in turn allows for a better comparison with experiments. Additionally, in the last part of this work we shall incorporate in the analysis the interaction with electromagnetic modes. This analysis reinforces the results obtained in \cite{PhysRevLett.116.161303,PhysRevD.95.026012} and prepares the ground for further investigations and applications of the model. 

The remainder of this paper is organised as follows. In Sec.\ref{II} we review the previous results and obtain the Hamiltonian formulation of the non-local oscillator. In Sec.\ref{III}, we find the ground state and the energy spectrum using standard time-independent perturbation theory. In Sec.\ref{IV}, we show that the previously found non-locality induced ``spontaneous time-periodic squeezing'' is preserved when more realistic initial states, with respect to the ones in~\cite{PhysRevLett.116.161303,PhysRevD.95.026012}, are considered. In Sec.\ref{V} the optomechanical interaction is introduced and the quantum Langevin's equations laid down. We conclude with Sec.\ref{VI} by presenting a summary and a discussion about possible future directions of this line of investigation.
\section{Non-local Schr\"odinger equation}\label{II}

In Refs.~\cite{PhysRevLett.116.161303,PhysRevD.95.026012}, the non-relativistic limit of an (analytic) non-local Klein--Gordon field was studied
, obtaining a non-local version of the standard Schr\"odinger equation of quantum mechanics. We briefly review here the approach 
(cf. Ref.~\cite{PhysRevD.95.026012} for further details). It should be noted that, the term non-local here refers to the presence in the theory of kinetic operators with infinitely many derivatives, see e.g.~\cite{barnaby2008dynamics}. The infinite number of derivatives is functional to the stability of the theory. Higher-derivative theories with finitely many derivatives would present instabilities related to the Ostrogardsky’s theorem~\cite{ostrogradski1850m}. This notion of non-locality has to be distinguished from the concept of \textit{nonlocality} as used in quantum mechanics, which refers to violations of Bell's inequalities~\cite{RevModPhys.86.419}.

Consider a complex scalar field theory described by the Lagrangian
\be
\mathcal{L} = \phi(x)^*f(\Box+\mu^2)\phi(x)+c.c.,
\label{lag}
\ee
where $\Box = c^{-2}\del^2_t -\nabla^2$ is the standard d'Alembertian operator, $\mu=mc/\hbar$, and $\phi(x)$ is the scalar field. We assume the function $f$ to be entire analytic with a single zero coinciding with the zero of the standard Klein--Gordon operator, in order to have a ghost-free theory. We can thus assume the power-series expansion $f(z)=\sum_n b_n z^n$. The non-relativistic limit of this theory leads to the non-local Schr\"odinger equation
\begin{equation}
f(\mathcal{S})\psi(t,{\bf x})=V({\bf x})\psi(t,{\bf x}).
\label{nls}
\end{equation}
Here $V({\bf x})$ is an external potential that we shall assume to be harmonic, we have decomposed the scalar field as $\phi(\chi)=e^{-im c^2 t/\hbar}\psi(t,{\bf x})$, and the non-local operator $f(\mathcal{S})$ is given by
\begin{equation}
f(\mathcal{S})\equiv\mathcal{S}+\sum_{n=2}^{\infty}b_{n}\left(-\frac{2m}{\hbar^{2}}\right)^{n-1}\mathcal{S}^{n},
\end{equation}
in terms of the standard Schr\"odinger operator  $\mathcal{S}=i\hbar\,\partial_t+\hbar^2{\bm \nabla}^2/2m$. 

Implicit in the definition of $f$ is a non-locality scale $l_k$. In order to make the non-locality scale explicit we set $b_n =l_k^{2n-2}a_n$, so that
\begin{equation}
f(\mathcal{S}) = \sum_{n=1}^\infty \left(-\frac{2m}{\hbar^{2}}\right)^{n-1} a_n l_k^{2n-2}\mathcal{S}^n.
\end{equation}
Clearly, the theory reduces to a local one for $l_k\to0$. Considering now a one-dimensional problem with $V(x)=m\omega^2 x^2/2$, and defining dimensionless variables 
$\hat{t}=\omega t$, $\hat{x}= \sqrt{\omega m/\hbar}\,x$, and  $\hat{\psi}=\gamma \psi$, 
where $\gamma$ has dimensions of ${\rm Length}^{-1/2}$ we arrive at the non-local Schr\"odinger equation
\begin{equation}\label{seqn}
\left(\hat{\mathcal{S}}-2a_2\epsilon\hat{\mathcal{S}}^2
+\sum_{n=3}^{\infty}a_{n}\epsilon^{n-1}(-2)^{n-1}\hat{\mathcal{S}}^{n}\right)\hat{\psi}=\frac{1}{2}\hat{x}^{2}\hat{\psi},
\end{equation}
where we have introduced the dimensionless parameter $\epsilon=m\omega l_k^2/\hbar$ that, for physically reasonable choices of the quantities involved, is small enough to be treated as a perturbative parameter in the problem at hand
\footnote{Note that, while we limit our analysis to leading order corrections, the inclusion of higher order terms would call for a careful treatment of the convergence of the perturbative series, see ,e.g.,~ \cite{zinn1996quantum,Kleinert:2001ax}.}. 

In Ref.~\cite{PhysRevLett.116.161303,PhysRevD.95.026012}, Eq.~\eqref{seqn} was solved perturbatively at order $\epsilon$, starting from the ansazt $\psi=\sum_n\epsilon^n\psi_n$ of the solution, and perturbing around semiclassical states of the local harmonic oscillator, also known as coherent states. The corresponding results for an initially prepared coherent state showcased the occurrence  of spontaneous time-dependent squeezing as well as non-harmonicity in the mean position of the oscillator. 
From now on, while always using the rescaled physical quantities introduced above, we shall drop the hats for easiness of notation.

\subsection{Hamiltonian re-formulation of the problem}
We now aim at providing a suitable Hamiltonian formulation of the model at hand. To this end, two distinct strategies can be followed. The first strategy starts from field Lagrangian, derives the $00$-component of the associated stress-energy tensor (see Appendix~\ref{SET}), and finally restrict the study to the single-particle sector. The second one consists in obtaining an evolution equation for the propagator arising from Eq.~\eqref{seqn}. Unfortunately, the first approach is technically very challenging in light of the intricacies related to the projection onto the single-particle sector. Thus, we shall follow the second approach.  

Consider the time-evolution operator for a state vector in the Schr\"odinger picture. As for the local counterpart, we assume that in the non-local theory the full time-evolution operator $U(t)$ can be written as the exponential of an Hamiltonian operator $H$ as $U(t)=\exp(-i H t)$. This is consistent with the assumption that the Hamiltonian is the generator of time translations. {
We thus use $\psi(t, x)=U(t)\psi(0, x)$ to recast Eq.~\eqref{seqn} in the form
\begin{equation}\label{eeo}
    i\partial_{t}U(t)=H_{0}U(t)+2a_2 \epsilon {\cal S}^2 U(t)+\mathcal{O}(\epsilon^2),
\end{equation}
where $H_0$ is the Hamiltonian of the quantum harmonic oscillator. 

We can now expand the unknown Hamiltonian $H$ as 
    $H=\sum_{n=0}^{\infty}\epsilon^n h_{n}$,
and use it in Eq.~\eqref{eeo}. By equating terms of the same order in $\epsilon$, we see that (at the relevant order in the perturbative parameter) $h_0=H_0$ and $h_1=\mathcal{H}$ with
\begin{equation}\label{hcorrection}
    \mathcal{H}=2a_2 \left(H_0^2-p^2 H_0+\frac{p^4}{4}\right),
\end{equation}
where $p$ is the momentum operator. This result is valid for a general position-dependent potential and not just for the harmonic one under scrutiny here. This method can be used to find the form of the higher order corrections $h_{n}$ to the local Hamiltonian. 

We thus have the (effective) Hamiltonian for the non-local oscillator to first order in the perturbative parameter $\epsilon$ as $H\simeq H_{0}+\epsilon\mathcal{H}$. Further simple manipulations give 
\be
\begin{aligned}\label{hamilt}
    \mathcal{H}&=\frac{a_2}{2}\left([x^2,p^2]+ x^4\right)
     = a_2 (\mathcal{H}_+ +\mathcal{H}_{-}),
\end{aligned}
\ee
where $\mathcal{H}_{+}={x^4}/{2}$ and $\mathcal{H}_-=2i xp+1$, which are Hermitian and anti-Hermitian respectively. It is interesting to note that, as ${\cal H}_-=i\{x,p\}$, by neglecting the anharmonic part of the potential $\mathcal{H}_{+}$, we are left with (a particular case of) the so-called Swanson Hamiltonian~\cite{swanson2004transition,
1751-8121-48-5-055301,1751-8121-40-34-015}. 

\subsection{Consistency with the non-local Schr\"odinger equation}
Before using the Hamiltonian \eqref{hamil} to study the dynamics of the system, we show the consistency of the Schr\"odinger equation based on the former with that given in Eq.~\eqref{seqn}. 
Our non-local perturbative Hamiltonian implies
\begin{equation}\label{effH}
    i\partial_t |\psi(t)\rangle = (H_0 +\epsilon \mathcal{H})|\psi(t)\rangle,
\end{equation}
where $\ket{\psi(t)}$ is the quantum mechanical state of the rescaled scalar field $\psi$. 
Let us assume, as done in Ref.~\cite{PhysRevD.95.026012}, that $|\psi\rangle=|\psi_0\rangle+\sum_{n=1}\epsilon^n |\psi_{n}\rangle$, which we insert in~\eqref{effH}. By equating terms order by order in $\epsilon$, we obtain
\begin{align}
    & i\partial_t |\psi_0(t)\rangle = H_0 |\psi_0(t)\rangle,\\
    & i\partial_t |\psi_1(t)\rangle = H_0 |\psi_1(t)\rangle + \mathcal{H}|\psi_0(t)\rangle\label{second}.
\end{align}
We should now show that Eq.~\eqref{second} is equivalent to $i\partial_t |\psi_1(t)\rangle = H_0 |\psi_1(t)\rangle + 2a_2 {\cal S}^2|\psi_0(t)\rangle$, obtained in Refs.~\cite{PhysRevLett.116.161303,PhysRevD.95.026012}. We do it for a generic (position-dependent) potential. Consider $ 2{\cal S}^2 \psi_0$ and the $0^\text{th}$ order (in the perturbative parameter) consistency condition found in~\cite{PhysRevD.95.026012} when looking for the solution $\psi$ to Eq.~\eqref{seqn}, which reads $({\cal S}-V)\psi_0=0$. We thus have
\begin{equation}
    2{\cal S}^2 \psi_0=2{\cal S}(V\psi_0)=2V^2\psi_0+V''\psi_0+2V'\psi_0',
\end{equation}
where $g'=\partial_x g$ for any quantity $g$.

We now need to show that this is equivalent to 
\begin{equation}
    \mathcal{H}\psi_0=2V^2\psi_0+[V,p^2]\psi_0.
\end{equation}
Since we are considering a position-dependent potential, we can use $[p,f(x)]=-if'(x)$. It is then immediate to see that 
\begin{equation}
    \mathcal{H}\psi_0=2V^2\psi_0+V''\psi_0+2V'\psi_0',
\end{equation}
thus proving the equivalence for every state $\psi_{0}$ that is a solution of the local Schr\"odinger equation, and any potential $V(x)$.

\section{Time-independent perturbation theory: perturbed ground state and energy spectrum}\label{III}
The Hamiltonian formulation presented in the previous Section paves the way for addressing several interesting questions. We shall start by obtaining the perturbed ground state of the Hamiltonian, and its energy. We shall do so by treating $\epsilon\mathcal{H}$ as a small perturbation, and using standard time-independent perturbation theory. 

In such a framework, the perturbed ground state is given by $|\Omega\rangle\approx |0\rangle+\epsilon|0^{(1)}\rangle$, where
\begin{equation}\label{pstate}
    |0^{(1)}\rangle=\sum_{m\neq 0}\frac{ \langle m|\mathcal{H}|0\rangle}{E_{0}-E_{m}} |m\rangle.
\end{equation}
Here, $|m\rangle$ is the $m^\text{th}$ eigenstate of the unperturbed Hamiltonian and $E_m$ is the corresponding eigenvalue, i.e. $E_{m}=(m+1/2)$. 
In order to compute the expression in Eq.~\eqref{pstate} we need the matrix elements
\begin{equation}
    \langle m\neq 0|\mathcal{H}_{+}|0\rangle=\begin{cases}
\frac{3}{2\sqrt{2}}\,\, \text{for $m=2$},\\
\frac{\sqrt{3}}{2\sqrt{2}}\,\, \text{for $m=4$},\\
0 \,\, \text{otherwise}
\end{cases}
\end{equation}
and
\begin{equation}
    \langle m\neq 0|\mathcal{H}_{-}|0\rangle
    =\begin{cases}
-\sqrt{2}\,\, \text{for $m=2$},\\
0 \,\, \text{otherwise}.
\end{cases}
\end{equation}
Note that the constant in $\mathcal{H}_-$ does not play any role owing to the orthogonality of the elements of  the set $\{|m\rangle\}$ and to the fact that we sum over all $m\neq 0$.

Using the above matrix elements we find the perturbed ground state
\begin{equation}\label{pertground}
   |\Omega\rangle\approx |0\rangle+\epsilon\left(\frac{1}{4\sqrt{2}}|2\rangle-\frac{\sqrt{3}}{8\sqrt{2}}|4\rangle\right),
\end{equation}
with associated energy $\mathcal{E}_{\Omega}=\langle\Omega|(H_0+\epsilon\mathcal{H})|\Omega\rangle$, given by
\begin{equation}
    \mathcal{E}_{\Omega}\approx\frac{1}{2}+\epsilon E^{(1)}_{0}=\frac{1}{2}+\epsilon\langle 0|\mathcal{H}|0\rangle=\frac{1}{2}+\frac{3}{8}\epsilon.
\end{equation}
In this derivation we have used that $\langle 0|\mathcal{H_+}|0\rangle=3/8$ and $\langle 0|\mathcal{H_-}|0\rangle=0$. 
Note that the perturbed energy is real despite the fact that the perturbation $\mathcal{H}$ has an anti-Hermitian component. 

We can extend this approach to a generic perturbed energy-eigenstate $\ket{n}$ of the system by writing it as $\sim |n\rangle+\epsilon|n^{(1)}\rangle$ with
\begin{equation}\label{pstate2}
    |n^{(1)}\rangle=\sum_{k\neq n}\frac{\langle k|\mathcal{H}|n\rangle}{E_{n}-E_{k}} |k\rangle
\end{equation}
and using the matrix elements
\begin{equation}\label{matrixe}
    \langle k|\mathcal{H}|n\rangle=\begin{cases}
\frac{1}{8}\sqrt{\frac{n!}{(n-4)!}}\,\, \text{for}~k=n-4,\\
\frac{1}{8}\sqrt{\frac{(n+4)!}{n!}}\,\, \text{for}~k=n+4,\\
\frac{1}{4}\sqrt{\frac{n!}{(n-2)!}}(3+2n)\,\, \text{for}~k=n-2,\\
\frac{1}{4}\sqrt{\frac{(n+2)!}{n!}}(2n-1)\,\, \text{for}~k=n+2,\\
\frac{3}{8}(1+2n+2n^2)\,\, \text{for}~k=n,\\
0 \,\, \text{otherwise.}
\end{cases}
\end{equation}
This gives 
\be
\begin{aligned}\label{pstate3}
    |n^{(1)}\rangle&= \frac{1}{32}\sqrt{\frac{(n+4)!}{n!}}|n+4\rangle+\frac{1}{8}\sqrt{\frac{(n+2)!}{n!}}(2n-1)|n+2\rangle\\
    & -\frac{1}{8}\sqrt{\frac{n!}{(n-2)!}}(3+2n)\Theta(n-2)|n-2\rangle\\
    &-\frac{1}{32}\sqrt{\frac{n!}{(n-4)!}}\Theta(n-4)|n-4\rangle,
\end{aligned}
\ee
where $\Theta(0)$ is the Heaviside function and we have taken $\Theta(0)=1$. The corresponding corrections to the energy eigenvalues are
\begin{equation}
  E_{n}^{(1)}=\langle n|\mathcal{H}|n\rangle=\frac{3}{8}(1+2n+2n^2),
\end{equation} 
so that 
$  \mathcal{E}_{n}\sim \frac{1}{2}+n+\epsilon \frac{3}{8}(1+2n+2n^2)$.
It should be noted that, as for the ground state, the anti-Hermitian part of $\mathcal{H}$ does not contribute to the energy eigenvalues. Thus, the energy eigenvalues are the same as the ones of an anharmonic oscillator with anharmonicity modulated by $\epsilon$.


\section{Dynamics of physical and unphysical initial states}\label{IV}
Having identified a suitable Hamiltonian formulation, we now consider the evolution of an initial state, a problem that we tackle using time-dependent perturbation theory. 

The evolution of states perturbatively close to solutions of the standard Schr\"odinger equation for a harmonic oscillator was already considered in Ref.~\cite{PhysRevLett.116.161303,PhysRevD.95.026012}. 
However, in these previous works the solutions of the local Schr\"odinger equation --- in particular ground and coherent states --- were chosen as initial conditions. Such a choice was motivated by the requirement of perturbing around known states of the local harmonic oscillator and, at the same time, by the necessity to choose arbitrary yet reasonable initial conditions. 
Unfortunately, such a choice is incompatible with the experimental protocols aimed at testing non-locality, and should thus be abandoned. Indeed, the Hamiltonian formulation of the model allows us to go beyond such bottleneck, and thus concretely advance towards phenomenological tests of quantum gravity induced non-locality. 

Let us then consider an optomechanical platform where the motion of a massive harmonic oscillator is driven and, to some extent, controlled by its radiation-pressure coupling to a light mode~\cite{aspelmeyer2014cavity}. Such an interaction can be effectively used to reduce the mean energy of the oscillator~\cite{Gigan2006}, thus cooling the system all the way down to energies comparable to the ground state of its Hamiltonian. If non-locality is present, the steady-state that will be approached by the oscillator is the ground state of the non-local Hamiltonian that was obtained (perturbatively) in Sec.~\ref{III}. It is then clear that the latter state is a well-motivated physical initial condition (in contrast to the ground state of a local oscillator) for the study of the subsequent dynamical evolution and comparison with potential experimental verifications.  

An analogous argument holds for coherent states. In a realistic protocol, like the one described in refs.~\cite{PhysRevLett.116.161303,PhysRevD.95.026012}, the mechanical oscillator is cooled to the ground state and then displaced by a laser. In light of the argument put forward above, this would result in the displacement of the the non-local ground state of the oscillator. 

We thus investigate the dynamics of both physical and unphysical initial conditions. The latter analysis is presented here to show that we can recover all the previous results reported in Ref.~\cite{PhysRevLett.116.161303,PhysRevD.95.026012} without invoking any ansazt on the solutions of the non-local Schr\"odinger equation. The former study, instead, will lead to previously unreported results, enabled by the newly proposed Hamiltonian formulation, which are more relevant for experimental purposes. 

Let us stress that we will show that in the more physical realistic scenario, the perturbed ground state is indeed a stationary state, consistently with it being the ground state of the Hamiltonian, and thus does not show any time-dependent quadrature variances. This is in contrast with what was reported in Ref.~\cite{PhysRevLett.116.161303,PhysRevD.95.026012}, where the initial state was assumed to be the unperturbed ground state. However, in the case in which the system is initially displaced from the ground state, we do find that the "smoking gun'' effect of time-periodic variances (periodic squeezing) remains a feature of the dynamics. Remarkably, the effect is even enhanced with respect to the previous studies, while the state remains a minimum-uncertainty one.

\subsection{Time-dependent perturbation theory}
We use time-dependent perturbation theory to solve the equation $i\partial_t{\ket{\psi (t)}}=\left(H_0 +\epsilon\mathcal{H}\right)\ket{\psi(t)}  $ perturbatively. We thus write
\begin{equation}
    |\psi(t)\rangle =\Sigma_n a_n(t) e^{-i E_n t} |n\rangle,
\end{equation}
where $\{|n\rangle\}$ are energy eigenstates of the unperturbed Hamiltonian with eigenvalues $E_n=(n+1/2)$, and $a_n(t)$ are time-dependent coefficients. The latter satisfy the following dynamical equations 
\begin{equation}
    \frac{d a_k(t)}{dt}=-i \epsilon  \Sigma_n \langle k|\mathcal{H}|n\rangle a_n(t) e^{i\left(E_k-E_n\right)t},
\end{equation}
whose solutions are
\begin{equation}\label{aksol}
    a_k(t)=a_k(0)-i \epsilon  \Sigma_n \langle k|\mathcal{H}|n\rangle\int _{0}^{t}dt' a_n( t') e^{it' (k-n)},
\end{equation}
where we used $E_k-E_n=k-n$. Expanding the coefficients as $a_n(t)=a_n^{(0)}(t)+\epsilon a_n^{(1)}(t)+\mathcal{O}(\epsilon^2)$ we find 
\begin{equation}\label{perteq}
    \begin{cases}
        a^{(0)}_k(t)=a_{k}^{(0)}(0),\\
        a_k^{(1)}(t)=a_{k}^{(1)}(0)-i\sum_n a_n^{(0)}(0)\int_0^t dt'\langle k|\mathcal{H}|n\rangle e^{i(k-n)t'}.
    \end{cases}
\end{equation}

In order to go further we must specify an initial state and study its dynamics. In the following, we report the results for the dynamics of the two initial states of interest mentioned above. Further details can be found in Appendix \ref{TimeDepPert}.

\subsection{Initial condition: ground state}
Let us first consider the case in which the initial preparation is the ground state of the oscillator. We consider both the perturbed and the unperturbed ground states even if, as already discussed, the former is a more realistic choice if non-locality is assumed to play a role in the dynamics of the system.

\subsubsection{Local Hamiltonian (unperturbed) ground state}
We start with the unperturbed ground state, for which $a_k(0)=a_k^{(0)}(0)=\delta_{k0}$ and $a_k^{(1)}(0)=0$, so that 
\begin{align}
    a_{k}^{(1)}(t)=-i \mathcal{H}_{k0} \int _{0}^{t}dt' e^{i k t'}=-i{\cal H}_{k0}
    \begin{cases}
    \dfrac{2e^{i k t/2}}{k}\sin(k t/2)~\text{for}~k\neq0,\\
    t~\text{for}~k=0,
    \end{cases}
\end{align}
where we have introduced the notation $\mathcal{H}_{kn} := \langle k|\mathcal{H}|n\rangle$. 
We see here the appearance of secular terms in $a_0(t)$ (cf. Appendix \ref{TimeDepPert}), in full analogy with previous results. Such terms are a typical feature in time-dependent perturbation methods, and must be discarded as unphysical. 

We can however avoid them by operating in a slightly different way: instead of expanding all the coefficients $a_k$ in $\epsilon$, we introduce the \emph{ansatz}
\begin{equation}\label{ansatz}
    \begin{cases}
        a_0(t)= e^{-i \epsilon \mathcal{H}_{00} t}\,+\,\mathcal{O}(\epsilon^2),\\
        a_k(t)=\epsilon a_{k}^{(1)}(t)\,+\,\mathcal{O}(\epsilon^2) \qquad k \ne 0.
    \end{cases}
\end{equation}
By substituting it in Eq.~\eqref{aksol}, we can show that the equation holds to first order in $\epsilon$ for $k=0$, while for $k \ne 0$ it gives
\begin{eqnarray}
a_k^{(1)}(t) =&& -i \mathcal{H}_{k0} \int _{0}^{t}dt' e^{i (k-\epsilon \mathcal{H}_{00})  t'}\nonumber\\
&&=-i \mathcal{H}_{k0}\frac{2e^{i (k-\epsilon \mathcal{H}_{00}) t/2}}{k-\epsilon \mathcal{H}_{00}}\sin\left(\frac{(k-\epsilon \mathcal{H}_{00}) t}{2}\right)
\end{eqnarray}
This result can be simplified by taking the lowest order in $\epsilon$, thus recovering the expression for $k\neq0$ obtained earlier. The state at time $t$ is then given by
\begin{equation}
|\psi(t)\rangle = e^{-it/2}|0\rangle+i \epsilon \frac{e^{-i 3 t/2}}{2\sqrt2}\left(\sin(t) |2\rangle-\frac{\sqrt{3}}{2}e^{-i t}\sin(2t) |4\rangle\right).
\end{equation}

\subsubsection{Non-local Hamiltonian (perturbed) ground state}
Consider now the perturbed ground state in Eq.~\eqref{pertground} as initial state, we have 
\begin{align}
    &a_0(0)=1,~~a_2(0)=\epsilon\frac{1}{4\sqrt{2}},~~a_3(0)=-\epsilon \frac{\sqrt{3}}{8\sqrt{2}},
\end{align}
so that not all $a_{n}^{(1)}(0)$ vanish, unlike for the previous case.

Once the perturbative evolved state is determined, we can focus on physically relevant quantities, such as the statistics of position and momentum of the oscillator. The calculations involved to achieve this task are provided in Appendix~\ref{TimeDepPert}. Here we report the main features of the corresponding results. 

While it is easy to see that the mean position and momentum vanish (as in~\cite{PhysRevLett.116.161303,PhysRevD.95.026012}), as in the previous case, the variances are time-independent. If we consider a generic quadrature $x(\theta)=({a}e^{-i\theta}+{a}^\dag e^{i\theta})/\sqrt{2}$ of the oscillator (here $a$ and $a^\dag$ are the annihilation and creation operators of the oscillator), we find
\begin{align}
     \rm{Var}(x(\theta))=\frac{1}{2}+\frac{\epsilon}{4}\cos(2\theta)+\mathcal{O}(\epsilon^2).
\end{align}
The phenomenology of the first and second moments of $x(\theta)$ is consistent with the fact that we are looking at a stationary state of the perturbed Hamiltonian, which is a feature that can be checked by using Eq.~\eqref{statet}. Note that the product of the variances for conjugate observables is, at order $\epsilon$, equal to $1/4$, i.e., the state is a minimum uncertainty state without isotropic variances. 

\subsection{Initial condition: displaced vacuum}
Now we consider the case in which an experimental protocol, like the one described in Sec.~\ref{III}, is implemented. In this case, after a cooling phase that should ideally bring the oscillator to its ground state, a laser pulse displaces its state in phase space. Previously, an initial coherent state was assumed to model this situation. However, as already discussed, a much more realistic assumption is to consider the displaced perturbed ground state.

In Appendix~\ref{TimeDepPert} we consider in detail the displaced ground state as an initial state and give the expressions to compute the mean value and variance of a generic quadrature. Here we simply report the results concerning the mean values and variances of position and momentum. Note that the perturbed states, to first order in $\epsilon$, have in general a time-dependent norm due to the non-Hermiticity of the perturbed Hamiltonian. As in~\cite{PhysRevLett.116.161303,PhysRevD.95.026012}, all the results below are obtained by normalizing the states with their time-dependent norm.

The mean values (see Fig.\ref{means}), are given by the same expression as for the initial standard coherent state in~\cite{PhysRevLett.116.161303,PhysRevD.95.026012}, i.e., 
\begin{align}
    & \langle x\rangle=\alpha \sqrt{2}  \cos (t)\,\Bigg\{1+\frac{1}{4}\epsilon\alpha^2[\cos(2t)-1]\Bigg\}\\
    & \langle p\rangle=-\alpha\sqrt{2} \sin (t)\,\Bigg\{1+\frac{1}{4}\epsilon[\alpha^2(7+3\cos(2t))-2]\Bigg\},
\end{align}
where $\alpha$ is the amplitude of the displacement (assumed to be real for easiness of analysis). The presence of a third harmonic in the mean position is clearly related to the Hermitian part of the Hamiltonian perturbation $\mathcal{H}$.
\begin{figure}[h]
\centering
\includegraphics[width=0.9\columnwidth]{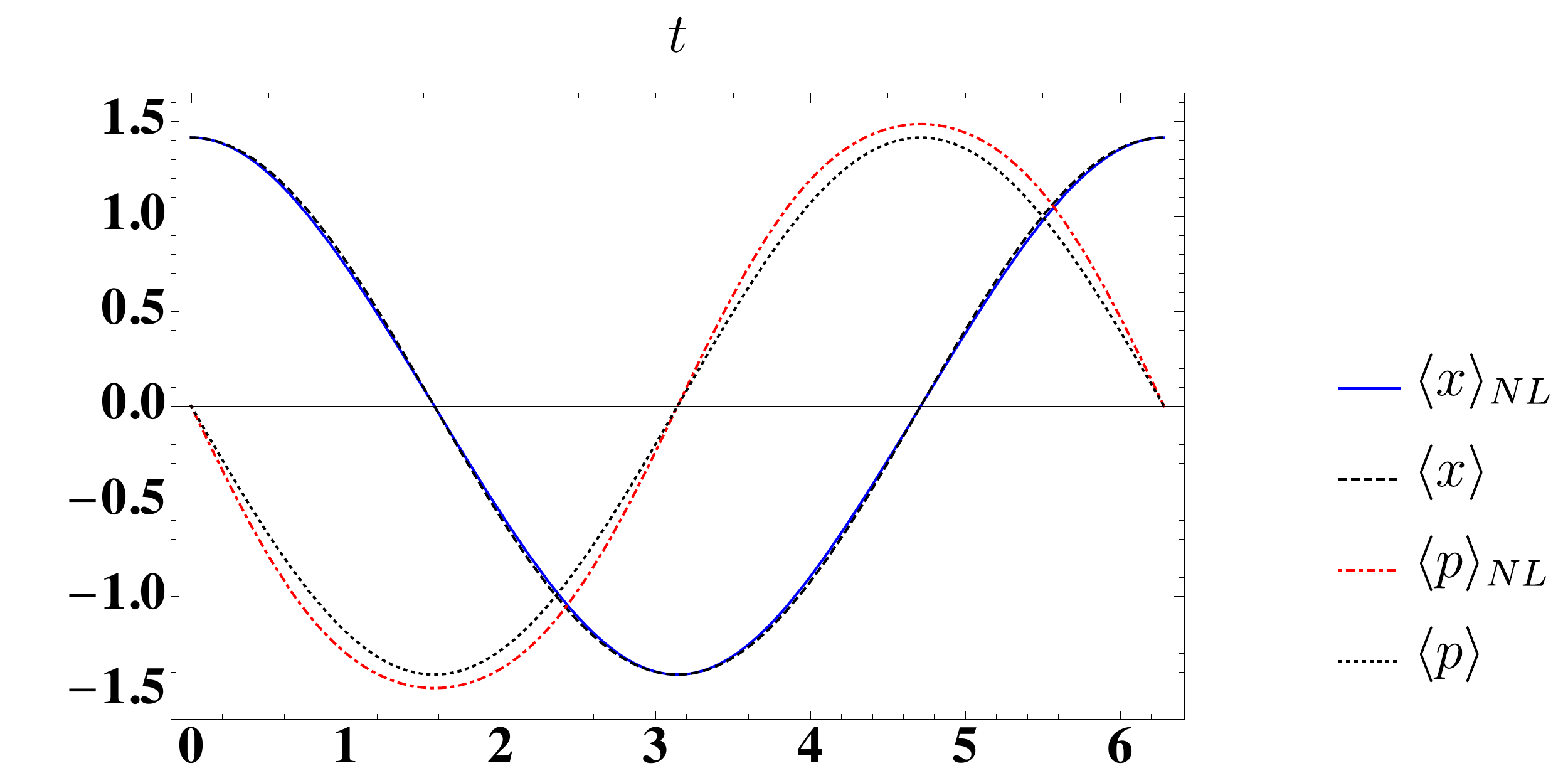}
  \caption{Time dependence of the mean position and momentum for a state initially in the displaced perturbed-ground state (colours online). We choose $\alpha=1$ and we have set $\epsilon=10^{-1}$ to magnify the effect of non-locality. The continuous blue and red lines represent the mean position and momentum for the non-local oscillator, while the black dotted and black dashed lines represent the mean position and momentum  for the local coherent state evolution}
  \label{means}
\end{figure}
The variances are given by
\begin{align}
    &\rm{Var}(x)=\frac{1}{2}+\epsilon  \left\{\frac{1}{4}-\frac{3}{2}\alpha^2\left[1-\cos (2 t)\right]\right\}\\
    &\rm{Var}(p)=\frac{1}{2}-\epsilon  \left\{\frac{1}{4}-\frac{3}{2}\alpha^2\left[1-\cos(2t)\right]\right\},
\end{align}
and differ from the case analyzed in~\cite{PhysRevLett.116.161303,PhysRevD.95.026012}. Nevertheless, it can be easily seen that their product is still equal to $1/4$ at order $\epsilon$. Thus we see that the modulation of the variances remains 
a robust characteristic of non-local dynamics when we displace the \textit{perturbed} ground state at the start of the experimental protocol. It should be noted that, in comparison with the modulation found in~\cite{PhysRevLett.116.161303,PhysRevD.95.026012} the variances acquire an extra constant shift of order $\epsilon$, which enhances the modulation effects. 
\begin{figure}[h]
\centering
\includegraphics[width=1.0\columnwidth]{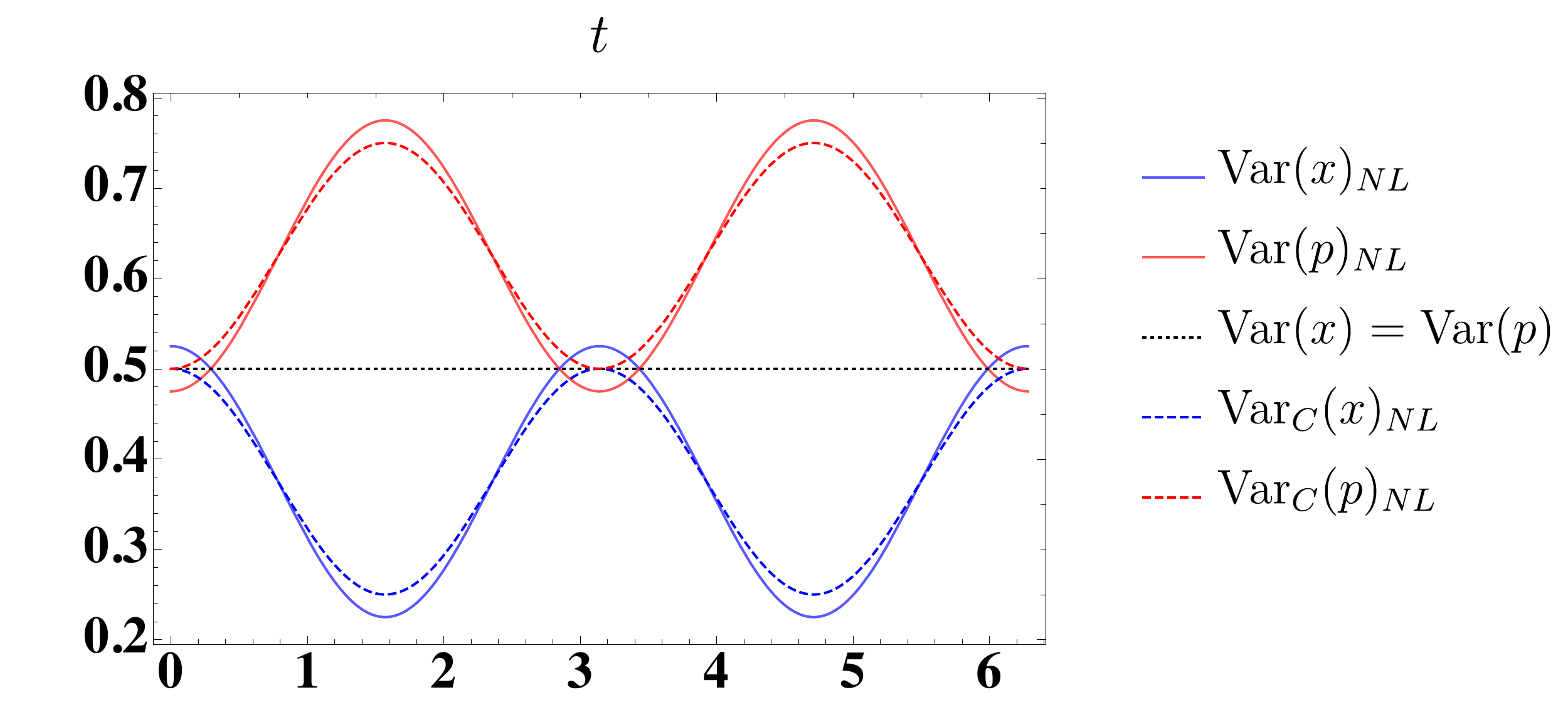}
  \caption{Periodic time dependence of the variances of position and momentum for a state initially in the displaced perturbed-ground state (color online). We choose $\alpha=1$ and we have set $\epsilon=10^{-1}$ to magnify the effect of non-locality. The continuous blue and red lines represent the variance of position and momentum respectively. The black dotted line represents the variance of position and momentum for the local coherent state which are equal to $1/2$. The dashed blue and red lines represent the variance of position and momentum, $\rm{Var}_C(x), \rm{Var}_C(p)$ respectively, when the initial state is a standard coherent state (i.e. when it is the displaced unperturbed ground state) like the one considered in~\cite{PhysRevLett.116.161303,PhysRevD.95.026012}. Note how the variances in the case of the more physical initial state (the displaced perturbed ground state) are enhanced with respect to previous findings.   
}
  \label{vars}
\end{figure}

\section{Interaction with electromagnetic modes}\label{V}
A great advantage of the Hamiltonian formulation, with respect to the original Schr\"odinger point of view of Refs.~\cite{PhysRevLett.116.161303,PhysRevD.95.026012}, is the possibility to generalize the description to include an interaction of the mechanical oscillator with the outer world, thus providing a more realistic description directly comparable to experiments. Two main ingredients are necessary to complete this picture, namely the interaction with a measurement system and dissipation. For the former, we refer to cavity optomechanics~\cite{aspelmeyer2014cavity} which have recently achieved the major breakthrough of preparing and observing macroscopic mechanical oscillators in non-classical states. The readout of the oscillator motion is provided by its mechanical interaction with an electromagnetic field (in the optical or microwave domain) inside a (cavity) resonator. The full Hamiltonian can be written as~\cite{law1995interaction,aspelmeyer2014cavity}
\begin{equation}
H_{om} = \hbar \omega H + \hbar\Delta f^{\dag} f - \hbar g_0 f^\dag f \,(b^\dag+b),
\end{equation}
where $\Delta$ is the detuning of the radiation frequency with respect to the cavity resonance, $f$ ($b$) is the cavity mode (mechanical mode)  annihilation operator (in our notation, $x = (b^\dag+b)/\sqrt{2}$). 
Note that the electromagnetic field is treated as if described by a local dynamics here. This is justified by the fact that the
massless, relativistic light field used in actual optomechanical experiments is a low-energy field and no relevant deviations from locality are expected in this case\footnote{A more precise dimensional argument goes as follows. First order non-local corrections to the d’Alembertian of the relativistic field go like $\Box/\Lambda^2$, where $\Lambda=\hbar c/\ell_k$ is the non-locality energy scale. For this to be subleading to the first order correction felt by the massive oscillator we must have that $E^2/\Lambda^2\ll\epsilon$, where $\epsilon=m\omega\ell_k^2/\hbar$, or, equivalently $\hbar \omega_{em}^2\ll\omega m c^2$. Assuming a very conservative frequency of $10^{15}$Hz for the electromagnetic field gives $m\omega\gg 10^{-12}$kg/s, which is easily satisfied by the optomechanical experiments of interest. Indeed, note that experiments aiming to explore possible non-local effects at the smallest possible scale should have the largest $m\omega$ product compatible with the achievement of the quantum regime, which, presently, is at least of the order of $10^{-5}$ kg/s}.

Under typical conditions of intense intracavity field, the Hamiltonian can be linearized with respect to the field operators, obtaining the quadratic form
\begin{equation}\label{Hom}
H'_{om} = \hbar  \omega H + \hbar \Delta f^{\dag} f - \hbar g (f^\dag + f) (b^\dag+b),
\end{equation}
where $g = \sqrt{n_c} \, g_0$ is the optomechanical coupling strength and $n_c$ is the mean intracavity photon number.   

The coupling of the optomechanical system to the environment is modelled within a Langevin approach, where the Heisenberg equations are complemented by quantum noise sources and dissipation terms. As ${\cal H}$ contains a non-Hermitian part, the Heisenberg equations for a generic operator $O$ are $\partial_t O=i[{\cal H}_{+},O]-i\{{\cal H}_{-},O\}$, and the Langevin equations for the open optomechanical system are
\be
\begin{aligned}
     \dot{f}&=-\left(i\Delta+\frac{\kappa}{2}\right)f+ig(b^\dag + b)+\sqrt{\kappa}f_{in}(t),\\
     \dot{b}&=-\left(i\omega+\frac{\Gamma}{2}\right)b+i g (f^\dag+f)+\sqrt{\Gamma}b_{in}(t)\\
    &+i\omega \epsilon \left[\frac{1}{2}(b+b^\dag)^3-(4b-2b^3+2b^\dag b b^{\dag}) \right],
\end{aligned}
\ee
where $\kappa$ is the photon output coupling rate, $\Gamma$ is the mechanical energy damping rate, $f_{in}$ is the field at the cavity input, and $b_{in}$ is the thermal noise affecting the motion of the oscillator. Measurements of the oscillator properties are actually performed by detecting the output field $f_{out}$, related to $f$ by the input-output relations $f_{out} = -f_{in}+\sqrt{\kappa} f$~\cite{WallsMilburn}.

For $\epsilon=0$ we recover the standard  quantum optomechanical Langevin model. It has been shown both theoretically and experimentally that the regime corresponding to the choice of detuning $\Delta \simeq -\omega$ allows for the preparation of the mechanical oscillator in states that are very close to the ground state~\cite{chan2011laser,safavi2012observation,teufel2011sideband}. For $\epsilon \ne 0$ anharmonic terms generate effects that can potentially be detected in optomechanics experiments, and whose detailed analysis deserves further investigations.

\section{Discussion and Conclusions}\label{VI}
In this work we have obtained a Hamiltonian formulation of the non-local harmonic oscillator introduced in Refs.~\cite{PhysRevLett.116.161303,PhysRevD.95.026012}. This allowed us to recover previously reported results using only standard perturbation theory, without having to explicitly solve any non-local differential equation or make an ansatz on the form of the solutions. 

Moreover, such formulation enabled the derivation of the ground state and energy spectrum of the Hamiltonian using time-independent perturbation theory, thus identifying physical initial preparations that can be used to study the effects of non-locality on the dynamical evolution of the system, an important cornerstone for comparison with experiments. 

In particular, the perturbed ground state and the displaced ground state have been considered as realistic initial states that can be reached by feasible experimental protocols. Indeed, by cooling the system the perturbed ground state should be approached, assuming non-locality is present, while displacing this state with a laser field in an optomechanical set-up would clearly prepare the system in the corresponding displaced state.

Studying the time evolution of physical preparation states, we showed that the perturbed ground state does not have, consistently, any relevant time dependence. Moreover, it does not present isotropic variances of conjugate quadratures --- in contrast to the local case ---, while remaining a minimum uncertainty state. The time evolution of an initially displaced ground state showed that the time-dependent state presents oscillating variances for conjugate observables such that it remains of minimum uncertainty. This reinforces the claims of Refs.~\cite{PhysRevLett.116.161303,PhysRevD.95.026012} where a similar result was obtained. 

Finally, we introduced the interaction between the non-local oscillator and a cavity electromagnetic mode and derived the quantum Langevin equations describing the system and dissipation terms. It is straightforward from here to obtain a master equation for the mechanical mode, or even reformulate the whole evolution in phase-space via Fokker--Planck type equations. We leave this for future investigations.

We believe that  this work paves the way to the study of the non-local oscillator model of Refs.~\cite{PhysRevLett.116.161303,PhysRevD.95.026012} in realistic conditions attainable by current and near-future optomechanical experiments. Future endeavours are to obtain solutions to the master equation, thus taking into account the effects of environmental decoherence on the dynamics of the non-local system. This would allow one to consider more general experimental protocols with respect to those that have been previously discussed. The goal will be to obtain predictions which can be compared with actual experiments in order to cast constraints on possible non-localities, or even rule out specific non-local models.


\acknowledgements
AB thanks Marco Letizia for useful discussions. 
AB is supported by H2020 through the MSCA IF pERFEcTO (Grant No. 795782). MP acknowledges support from the DfE-SFI Investigator Programme (Grant No. 15/IA/2864), the H2020 Collaborative Project TEQ (Grant Agreement No. 766900), and the Leverhulme Trust Research Project Grant.


\bibliography{references2.bib}

\begin{thebibliography}{38}%
\makeatletter
\providecommand \@ifxundefined [1]{%
 \@ifx{#1\undefined}
}%
\providecommand \@ifnum [1]{%
 \ifnum #1\expandafter \@firstoftwo
 \else \expandafter \@secondoftwo
 \fi
}%
\providecommand \@ifx [1]{%
 \ifx #1\expandafter \@firstoftwo
 \else \expandafter \@secondoftwo
 \fi
}%
\providecommand \natexlab [1]{#1}%
\providecommand \enquote  [1]{``#1''}%
\providecommand \bibnamefont  [1]{#1}%
\providecommand \bibfnamefont [1]{#1}%
\providecommand \citenamefont [1]{#1}%
\providecommand \href@noop [0]{\@secondoftwo}%
\providecommand \href [0]{\begingroup \@sanitize@url \@href}%
\providecommand \@href[1]{\@@startlink{#1}\@@href}%
\providecommand \@@href[1]{\endgroup#1\@@endlink}%
\providecommand \@sanitize@url [0]{\catcode `\\12\catcode `\$12\catcode
  `\&12\catcode `\#12\catcode `\^12\catcode `\_12\catcode `\%12\relax}%
\providecommand \@@startlink[1]{}%
\providecommand \@@endlink[0]{}%
\providecommand \url  [0]{\begingroup\@sanitize@url \@url }%
\providecommand \@url [1]{\endgroup\@href {#1}{\urlprefix }}%
\providecommand \urlprefix  [0]{URL }%
\providecommand \Eprint [0]{\href }%
\providecommand \doibase [0]{http://dx.doi.org/}%
\providecommand \selectlanguage [0]{\@gobble}%
\providecommand \bibinfo  [0]{\@secondoftwo}%
\providecommand \bibfield  [0]{\@secondoftwo}%
\providecommand \translation [1]{[#1]}%
\providecommand \BibitemOpen [0]{}%
\providecommand \bibitemStop [0]{}%
\providecommand \bibitemNoStop [0]{.\EOS\space}%
\providecommand \EOS [0]{\spacefactor3000\relax}%
\providecommand \BibitemShut  [1]{\csname bibitem#1\endcsname}%
\let\auto@bib@innerbib\@empty
\bibitem [{\citenamefont {Hossenfelder}(2010)}]{hossenfelder2010experimental}%
  \BibitemOpen
  \bibfield  {author} {\bibinfo {author} {\bibfnamefont {Sabine}\ \bibnamefont
  {Hossenfelder}},\ }\bibfield  {title} {\enquote {\bibinfo {title}
  {Experimental search for quantum gravity},}\ }in\ \href@noop {} {\emph
  {\bibinfo {booktitle} {Workshop on Experimental Search for Quantum Gravity
  NORDITA, Stockholm, Sweden}}}\ (\bibinfo {organization} {Springer},\ \bibinfo
  {year} {2010})\BibitemShut {NoStop}%
\bibitem [{\citenamefont {Liberati}\ and\ \citenamefont
  {Maccione}(2011)}]{liberati2011quantum}%
  \BibitemOpen
  \bibfield  {author} {\bibinfo {author} {\bibfnamefont {Stefano}\ \bibnamefont
  {Liberati}}\ and\ \bibinfo {author} {\bibfnamefont {Luca}\ \bibnamefont
  {Maccione}},\ }\bibfield  {title} {\enquote {\bibinfo {title} {Quantum
  gravity phenomenology: achievements and challenges},}\ }in\ \href@noop {}
  {\emph {\bibinfo {booktitle} {Journal of Physics: Conference Series}}},\
  Vol.\ \bibinfo {volume} {314}\ (\bibinfo {organization} {IOP Publishing},\
  \bibinfo {year} {2011})\ p.\ \bibinfo {pages} {012007}\BibitemShut {NoStop}%
\bibitem [{\citenamefont {Amelino-Camelia}(2002)}]{AmelinoCamelia:2002vw}%
  \BibitemOpen
  \bibfield  {author} {\bibinfo {author} {\bibfnamefont {Giovanni}\
  \bibnamefont {Amelino-Camelia}},\ }\bibfield  {title} {\enquote {\bibinfo
  {title} {{Quantum gravity phenomenology: Status and prospects}},}\ }\bibfield
   {booktitle} {\emph {\bibinfo {booktitle} {{The interface of gravitational
  and quantum realms. Proceedings, 1st Inter-University Centre for Astronomy
  and Astrophysics Meeting, Pune, India, December 17-21, 2001}}},\ }\href
  {\doibase 10.1142/S0217732302007612} {\bibfield  {journal} {\bibinfo
  {journal} {Mod. Phys. Lett.}\ }\textbf {\bibinfo {volume} {A17}},\ \bibinfo
  {pages} {899--922} (\bibinfo {year} {2002})},\ \Eprint
  {http://arxiv.org/abs/gr-qc/0204051} {arXiv:gr-qc/0204051 [gr-qc]}
  \BibitemShut {NoStop}%
\bibitem [{\citenamefont {Liberati}(2013)}]{Liberati:2013xla}%
  \BibitemOpen
  \bibfield  {author} {\bibinfo {author} {\bibfnamefont {Stefano}\ \bibnamefont
  {Liberati}},\ }\bibfield  {title} {\enquote {\bibinfo {title} {{Tests of
  Lorentz invariance: a 2013 update}},}\ }\href {\doibase
  10.1088/0264-9381/30/13/133001} {\bibfield  {journal} {\bibinfo  {journal}
  {Class. Quant. Grav.}\ }\textbf {\bibinfo {volume} {30}},\ \bibinfo {pages}
  {133001} (\bibinfo {year} {2013})},\ \Eprint {http://arxiv.org/abs/1304.5795}
  {arXiv:1304.5795 [gr-qc]} \BibitemShut {NoStop}%
\bibitem [{\citenamefont {Belenchia}\ \emph
  {et~al.}(2016{\natexlab{a}})\citenamefont {Belenchia}, \citenamefont
  {Benincasa}, \citenamefont {Liberati}, \citenamefont {Marin}, \citenamefont
  {Marino},\ and\ \citenamefont {Ortolan}}]{PhysRevLett.116.161303}%
  \BibitemOpen
  \bibfield  {author} {\bibinfo {author} {\bibfnamefont {Alessio}\ \bibnamefont
  {Belenchia}}, \bibinfo {author} {\bibfnamefont {Dionigi M.~T.}\ \bibnamefont
  {Benincasa}}, \bibinfo {author} {\bibfnamefont {Stefano}\ \bibnamefont
  {Liberati}}, \bibinfo {author} {\bibfnamefont {Francesco}\ \bibnamefont
  {Marin}}, \bibinfo {author} {\bibfnamefont {Francesco}\ \bibnamefont
  {Marino}}, \ and\ \bibinfo {author} {\bibfnamefont {Antonello}\ \bibnamefont
  {Ortolan}},\ }\bibfield  {title} {\enquote {\bibinfo {title} {Testing quantum
  gravity induced nonlocality via optomechanical quantum oscillators},}\ }\href
  {\doibase 10.1103/PhysRevLett.116.161303} {\bibfield  {journal} {\bibinfo
  {journal} {Phys. Rev. Lett.}\ }\textbf {\bibinfo {volume} {116}},\ \bibinfo
  {pages} {161303} (\bibinfo {year} {2016}{\natexlab{a}})}\BibitemShut
  {NoStop}%
\bibitem [{\citenamefont {Belenchia}\ \emph
  {et~al.}(2016{\natexlab{b}})\citenamefont {Belenchia}, \citenamefont
  {Benincasa}, \citenamefont {Martin-Martinez},\ and\ \citenamefont
  {Saravani}}]{Belenchia:2016sym}%
  \BibitemOpen
  \bibfield  {author} {\bibinfo {author} {\bibfnamefont {Alessio}\ \bibnamefont
  {Belenchia}}, \bibinfo {author} {\bibfnamefont {Dionigi M.~T.}\ \bibnamefont
  {Benincasa}}, \bibinfo {author} {\bibfnamefont {Eduardo}\ \bibnamefont
  {Martin-Martinez}}, \ and\ \bibinfo {author} {\bibfnamefont {Mehdi}\
  \bibnamefont {Saravani}},\ }\bibfield  {title} {\enquote {\bibinfo {title}
  {{Low energy signatures of nonlocal field theories}},}\ }\href {\doibase
  10.1103/PhysRevD.94.061902} {\bibfield  {journal} {\bibinfo  {journal} {Phys.
  Rev.}\ }\textbf {\bibinfo {volume} {D94}},\ \bibinfo {pages} {061902}
  (\bibinfo {year} {2016}{\natexlab{b}})},\ \Eprint
  {http://arxiv.org/abs/1605.03973} {arXiv:1605.03973 [quant-ph]} \BibitemShut
  {NoStop}%
\bibitem [{\citenamefont {Sorkin}(2009)}]{sorkin2009does}%
  \BibitemOpen
  \bibfield  {author} {\bibinfo {author} {\bibfnamefont {Rafael~D}\
  \bibnamefont {Sorkin}},\ }\bibfield  {title} {\enquote {\bibinfo {title}
  {Does locality fail at intermediate length-scales},}\ }\href@noop {}
  {\bibfield  {journal} {\bibinfo  {journal} {Approaches to Quantum Gravity,
  Editor D. Oriti, Cambridge University Press, Cambridge}\ ,\ \bibinfo {pages}
  {26--43}} (\bibinfo {year} {2009})}\BibitemShut {NoStop}%
\bibitem [{\citenamefont {Barnaby}\ and\ \citenamefont
  {Kamran}(2008)}]{barnaby2008dynamics}%
  \BibitemOpen
  \bibfield  {author} {\bibinfo {author} {\bibfnamefont {Neil}\ \bibnamefont
  {Barnaby}}\ and\ \bibinfo {author} {\bibfnamefont {Niky}\ \bibnamefont
  {Kamran}},\ }\bibfield  {title} {\enquote {\bibinfo {title} {Dynamics with
  infinitely many derivatives: the initial value problem},}\ }\href
  {http://stacks.iop.org/1126-6708/2008/i=02/a=008} {\bibfield  {journal}
  {\bibinfo  {journal} {Journal of High Energy Physics}\ }\textbf {\bibinfo
  {volume} {2008}},\ \bibinfo {pages} {008} (\bibinfo {year}
  {2008})}\BibitemShut {NoStop}%
\bibitem [{\citenamefont {Biswas}\ and\ \citenamefont
  {Okada}(2015)}]{Biswas:2014yia}%
  \BibitemOpen
  \bibfield  {author} {\bibinfo {author} {\bibfnamefont {Tirthabir}\
  \bibnamefont {Biswas}}\ and\ \bibinfo {author} {\bibfnamefont {Nobuchika}\
  \bibnamefont {Okada}},\ }\bibfield  {title} {\enquote {\bibinfo {title}
  {{Towards LHC physics with nonlocal Standard Model}},}\ }\href {\doibase
  10.1016/j.nuclphysb.2015.06.023} {\bibfield  {journal} {\bibinfo  {journal}
  {Nucl. Phys.}\ }\textbf {\bibinfo {volume} {B898}},\ \bibinfo {pages}
  {113--131} (\bibinfo {year} {2015})},\ \Eprint
  {http://arxiv.org/abs/1407.3331} {arXiv:1407.3331 [hep-ph]} \BibitemShut
  {NoStop}%
\bibitem [{\citenamefont {Gambini}\ and\ \citenamefont
  {Pullin}(2014)}]{Gambini:2014kba}%
  \BibitemOpen
  \bibfield  {author} {\bibinfo {author} {\bibfnamefont {Rodolfo}\ \bibnamefont
  {Gambini}}\ and\ \bibinfo {author} {\bibfnamefont {Jorge}\ \bibnamefont
  {Pullin}},\ }\bibfield  {title} {\enquote {\bibinfo {title} {{Emergence of
  stringlike physics from Lorentz invariance in loop quantum gravity}},}\
  }\href {\doibase 10.1142/S0218271814420231} {\bibfield  {journal} {\bibinfo
  {journal} {Int. J. Mod. Phys.}\ }\textbf {\bibinfo {volume} {D23}},\ \bibinfo
  {pages} {1442023} (\bibinfo {year} {2014})},\ \Eprint
  {http://arxiv.org/abs/1406.2610} {arXiv:1406.2610 [gr-qc]} \BibitemShut
  {NoStop}%
\bibitem [{\citenamefont {Saravani}\ and\ \citenamefont
  {Aslanbeigi}(2015)}]{Saravani:2015rva}%
  \BibitemOpen
  \bibfield  {author} {\bibinfo {author} {\bibfnamefont {Mehdi}\ \bibnamefont
  {Saravani}}\ and\ \bibinfo {author} {\bibfnamefont {Siavash}\ \bibnamefont
  {Aslanbeigi}},\ }\bibfield  {title} {\enquote {\bibinfo {title} {{Dark Matter
  From Spacetime Nonlocality}},}\ }\href {\doibase 10.1103/PhysRevD.92.103504}
  {\bibfield  {journal} {\bibinfo  {journal} {Phys. Rev.}\ }\textbf {\bibinfo
  {volume} {D92}},\ \bibinfo {pages} {103504} (\bibinfo {year} {2015})},\
  \Eprint {http://arxiv.org/abs/1502.01655} {arXiv:1502.01655 [hep-th]}
  \BibitemShut {NoStop}%
\bibitem [{\citenamefont {Bojowald}\ \emph {et~al.}(2005)\citenamefont
  {Bojowald}, \citenamefont {Morales-T\'ecotl},\ and\ \citenamefont
  {Sahlmann}}]{PhysRevD.71.084012}%
  \BibitemOpen
  \bibfield  {author} {\bibinfo {author} {\bibfnamefont {Martin}\ \bibnamefont
  {Bojowald}}, \bibinfo {author} {\bibfnamefont {Hugo~A.}\ \bibnamefont
  {Morales-T\'ecotl}}, \ and\ \bibinfo {author} {\bibfnamefont {Hanno}\
  \bibnamefont {Sahlmann}},\ }\bibfield  {title} {\enquote {\bibinfo {title}
  {Loop quantum gravity phenomenology and the issue of lorentz invariance},}\
  }\href {\doibase 10.1103/PhysRevD.71.084012} {\bibfield  {journal} {\bibinfo
  {journal} {Phys. Rev. D}\ }\textbf {\bibinfo {volume} {71}},\ \bibinfo
  {pages} {084012} (\bibinfo {year} {2005})}\BibitemShut {NoStop}%
\bibitem [{\citenamefont {Belenchia}\ \emph {et~al.}(2015)\citenamefont
  {Belenchia}, \citenamefont {Benincasa},\ and\ \citenamefont
  {Liberati}}]{Belenchia:2014fda}%
  \BibitemOpen
  \bibfield  {author} {\bibinfo {author} {\bibfnamefont {Alessio}\ \bibnamefont
  {Belenchia}}, \bibinfo {author} {\bibfnamefont {Dionigi M.~T.}\ \bibnamefont
  {Benincasa}}, \ and\ \bibinfo {author} {\bibfnamefont {Stefano}\ \bibnamefont
  {Liberati}},\ }\bibfield  {title} {\enquote {\bibinfo {title} {{Nonlocal
  Scalar Quantum Field Theory from Causal Sets}},}\ }\href {\doibase
  10.1007/JHEP03(2015)036} {\bibfield  {journal} {\bibinfo  {journal} {JHEP}\
  }\textbf {\bibinfo {volume} {03}},\ \bibinfo {pages} {036} (\bibinfo {year}
  {2015})},\ \Eprint {http://arxiv.org/abs/1411.6513} {arXiv:1411.6513 [gr-qc]}
  \BibitemShut {NoStop}%
\bibitem [{\citenamefont {Hossenfelder}(2008)}]{Hossenfelder:2007re}%
  \BibitemOpen
  \bibfield  {author} {\bibinfo {author} {\bibfnamefont {S.}~\bibnamefont
  {Hossenfelder}},\ }\bibfield  {title} {\enquote {\bibinfo {title} {{A Note on
  Quantum Field Theories with a Minimal Length Scale}},}\ }\href {\doibase
  10.1088/0264-9381/25/3/038003} {\bibfield  {journal} {\bibinfo  {journal}
  {Class. Quant. Grav.}\ }\textbf {\bibinfo {volume} {25}},\ \bibinfo {pages}
  {038003} (\bibinfo {year} {2008})},\ \Eprint {http://arxiv.org/abs/0712.2811}
  {arXiv:0712.2811 [hep-th]} \BibitemShut {NoStop}%
\bibitem [{\citenamefont {Modesto}\ and\ \citenamefont
  {Rachwał}(2017)}]{Modesto:2017sdr}%
  \BibitemOpen
  \bibfield  {author} {\bibinfo {author} {\bibfnamefont {Leonardo}\
  \bibnamefont {Modesto}}\ and\ \bibinfo {author} {\bibfnamefont {Lesław}\
  \bibnamefont {Rachwał}},\ }\bibfield  {title} {\enquote {\bibinfo {title}
  {{Nonlocal quantum gravity: A review}},}\ }\href {\doibase
  10.1142/S0218271817300208} {\bibfield  {journal} {\bibinfo  {journal} {Int.
  J. Mod. Phys.}\ }\textbf {\bibinfo {volume} {D26}},\ \bibinfo {pages}
  {1730020} (\bibinfo {year} {2017})}\BibitemShut {NoStop}%
\bibitem [{\citenamefont {Barci}\ \emph {et~al.}(1996)\citenamefont {Barci},
  \citenamefont {Oxman},\ and\ \citenamefont {Rocca}}]{Barci:1995ad}%
  \BibitemOpen
  \bibfield  {author} {\bibinfo {author} {\bibfnamefont {D.~G.}\ \bibnamefont
  {Barci}}, \bibinfo {author} {\bibfnamefont {L.~E.}\ \bibnamefont {Oxman}}, \
  and\ \bibinfo {author} {\bibfnamefont {M.}~\bibnamefont {Rocca}},\ }\bibfield
   {title} {\enquote {\bibinfo {title} {{Canonical quantization of nonlocal
  field equations}},}\ }\href {\doibase 10.1142/S0217751X96001061} {\bibfield
  {journal} {\bibinfo  {journal} {Int. J. Mod. Phys.}\ }\textbf {\bibinfo
  {volume} {A11}},\ \bibinfo {pages} {2111--2126} (\bibinfo {year} {1996})},\
  \Eprint {http://arxiv.org/abs/hep-th/9503101} {arXiv:hep-th/9503101 [hep-th]}
  \BibitemShut {NoStop}%
\bibitem [{\citenamefont {Arzano}\ and\ \citenamefont
  {Consoli}(2018)}]{Arzano:2018gii}%
  \BibitemOpen
  \bibfield  {author} {\bibinfo {author} {\bibfnamefont {Michele}\ \bibnamefont
  {Arzano}}\ and\ \bibinfo {author} {\bibfnamefont {Luca~Tiberio}\ \bibnamefont
  {Consoli}},\ }\bibfield  {title} {\enquote {\bibinfo {title} {{Signal
  propagation on $\kappa$-Minkowski spacetime and nonlocal two-point
  functions}},}\ }\href {\doibase 10.1103/PhysRevD.98.106018} {\bibfield
  {journal} {\bibinfo  {journal} {Phys. Rev.}\ }\textbf {\bibinfo {volume}
  {D98}},\ \bibinfo {pages} {106018} (\bibinfo {year} {2018})},\ \Eprint
  {http://arxiv.org/abs/1808.02241} {arXiv:1808.02241 [hep-th]} \BibitemShut
  {NoStop}%
\bibitem [{\citenamefont {Tomboulis}(2015)}]{Tomboulis:2015gfa}%
  \BibitemOpen
  \bibfield  {author} {\bibinfo {author} {\bibfnamefont {E.~T.}\ \bibnamefont
  {Tomboulis}},\ }\bibfield  {title} {\enquote {\bibinfo {title} {{Nonlocal and
  quasilocal field theories}},}\ }\href {\doibase 10.1103/PhysRevD.92.125037}
  {\bibfield  {journal} {\bibinfo  {journal} {Phys. Rev.}\ }\textbf {\bibinfo
  {volume} {D92}},\ \bibinfo {pages} {125037} (\bibinfo {year} {2015})},\
  \Eprint {http://arxiv.org/abs/1507.00981} {arXiv:1507.00981 [hep-th]}
  \BibitemShut {NoStop}%
\bibitem [{\citenamefont {Taylor}\ and\ \citenamefont
  {Zwiebach}(2003)}]{Taylor:2003gn}%
  \BibitemOpen
  \bibfield  {author} {\bibinfo {author} {\bibfnamefont {Washington}\
  \bibnamefont {Taylor}}\ and\ \bibinfo {author} {\bibfnamefont {Barton}\
  \bibnamefont {Zwiebach}},\ }\bibfield  {title} {\enquote {\bibinfo {title}
  {{D-branes, tachyons, and string field theory}},}\ }in\ \href {\doibase
  10.1142/9789812702821_0012} {\emph {\bibinfo {booktitle} {{Strings, Branes
  and Extra Dimensions: TASI 2001: Proceedings}}}}\ (\bibinfo {year} {2003})\
  pp.\ \bibinfo {pages} {641--759},\ \Eprint
  {http://arxiv.org/abs/hep-th/0311017} {arXiv:hep-th/0311017 [hep-th]}
  \BibitemShut {NoStop}%
\bibitem [{\citenamefont {Szabo}(2003)}]{Szabo:2001kg}%
  \BibitemOpen
  \bibfield  {author} {\bibinfo {author} {\bibfnamefont {Richard~J.}\
  \bibnamefont {Szabo}},\ }\bibfield  {title} {\enquote {\bibinfo {title}
  {{Quantum field theory on noncommutative spaces}},}\ }\bibfield  {booktitle}
  {\emph {\bibinfo {booktitle} {{Frontiers of Mathematical Physics: Summer
  Workshop on Particles, Fields and Strings Burnaby, Canada, July 16-27,
  2001}}},\ }\href {\doibase 10.1016/S0370-1573(03)00059-0} {\bibfield
  {journal} {\bibinfo  {journal} {Phys. Rept.}\ }\textbf {\bibinfo {volume}
  {378}},\ \bibinfo {pages} {207--299} (\bibinfo {year} {2003})},\ \Eprint
  {http://arxiv.org/abs/hep-th/0109162} {arXiv:hep-th/0109162 [hep-th]}
  \BibitemShut {NoStop}%
\bibitem [{\citenamefont {Ostrogradsky}(1850)}]{ostrogradsky1850memoires}%
  \BibitemOpen
  \bibfield  {author} {\bibinfo {author} {\bibfnamefont {Michael}\ \bibnamefont
  {Ostrogradsky}},\ }\bibfield  {title} {\enquote {\bibinfo {title}
  {M{\'e}moires sur les {\'e}quations diff{\'e}rentielles, relatives au
  probl{\`e}me des isop{\'e}rim{\`e}tres},}\ }\href@noop {} {\bibfield
  {journal} {\bibinfo  {journal} {Mem. Acad. St. Petersbourg}\ }\textbf
  {\bibinfo {volume} {6}},\ \bibinfo {pages} {385--517} (\bibinfo {year}
  {1850})}\BibitemShut {NoStop}%
\bibitem [{\citenamefont {Belenchia}\ \emph {et~al.}(2017)\citenamefont
  {Belenchia}, \citenamefont {Benincasa}, \citenamefont {Liberati},
  \citenamefont {Marin}, \citenamefont {Marino},\ and\ \citenamefont
  {Ortolan}}]{PhysRevD.95.026012}%
  \BibitemOpen
  \bibfield  {author} {\bibinfo {author} {\bibfnamefont {Alessio}\ \bibnamefont
  {Belenchia}}, \bibinfo {author} {\bibfnamefont {Dionigi M.~T.}\ \bibnamefont
  {Benincasa}}, \bibinfo {author} {\bibfnamefont {Stefano}\ \bibnamefont
  {Liberati}}, \bibinfo {author} {\bibfnamefont {Francesco}\ \bibnamefont
  {Marin}}, \bibinfo {author} {\bibfnamefont {Francesco}\ \bibnamefont
  {Marino}}, \ and\ \bibinfo {author} {\bibfnamefont {Antonello}\ \bibnamefont
  {Ortolan}},\ }\bibfield  {title} {\enquote {\bibinfo {title} {Tests of
  quantum-gravity-induced nonlocality via optomechanical experiments},}\ }\href
  {\doibase 10.1103/PhysRevD.95.026012} {\bibfield  {journal} {\bibinfo
  {journal} {Phys. Rev. D}\ }\textbf {\bibinfo {volume} {95}},\ \bibinfo
  {pages} {026012} (\bibinfo {year} {2017})}\BibitemShut {NoStop}%
\bibitem [{\citenamefont {Ostrogradski}(1850)}]{ostrogradski1850m}%
  \BibitemOpen
  \bibfield  {author} {\bibinfo {author} {\bibfnamefont {M}~\bibnamefont
  {Ostrogradski}},\ }\bibfield  {title} {\enquote {\bibinfo {title} {M.
  ostrogradski, petersbourg 1, 18502 (1850).}}\ }\href@noop {} {\bibfield
  {journal} {\bibinfo  {journal} {Petersbourg}\ }\textbf {\bibinfo {volume}
  {1}},\ \bibinfo {pages} {18502} (\bibinfo {year} {1850})}\BibitemShut
  {NoStop}%
\bibitem [{\citenamefont {Brunner}\ \emph {et~al.}(2014)\citenamefont
  {Brunner}, \citenamefont {Cavalcanti}, \citenamefont {Pironio}, \citenamefont
  {Scarani},\ and\ \citenamefont {Wehner}}]{RevModPhys.86.419}%
  \BibitemOpen
  \bibfield  {author} {\bibinfo {author} {\bibfnamefont {Nicolas}\ \bibnamefont
  {Brunner}}, \bibinfo {author} {\bibfnamefont {Daniel}\ \bibnamefont
  {Cavalcanti}}, \bibinfo {author} {\bibfnamefont {Stefano}\ \bibnamefont
  {Pironio}}, \bibinfo {author} {\bibfnamefont {Valerio}\ \bibnamefont
  {Scarani}}, \ and\ \bibinfo {author} {\bibfnamefont {Stephanie}\ \bibnamefont
  {Wehner}},\ }\bibfield  {title} {\enquote {\bibinfo {title} {Bell
  nonlocality},}\ }\href {\doibase 10.1103/RevModPhys.86.419} {\bibfield
  {journal} {\bibinfo  {journal} {Rev. Mod. Phys.}\ }\textbf {\bibinfo {volume}
  {86}},\ \bibinfo {pages} {419--478} (\bibinfo {year} {2014})}\BibitemShut
  {NoStop}%
\bibitem [{\citenamefont {Zinn-Justin}(1996)}]{zinn1996quantum}%
  \BibitemOpen
  \bibfield  {author} {\bibinfo {author} {\bibfnamefont {Jean}\ \bibnamefont
  {Zinn-Justin}},\ }\href@noop {} {\emph {\bibinfo {title} {Quantum field
  theory and critical phenomena}}}\ (\bibinfo  {publisher} {Clarendon Press},\
  \bibinfo {year} {1996})\BibitemShut {NoStop}%
\bibitem [{\citenamefont {Kleinert}\ and\ \citenamefont
  {Schulte-Frohlinde}(2001)}]{Kleinert:2001ax}%
  \BibitemOpen
  \bibfield  {author} {\bibinfo {author} {\bibfnamefont {H.}~\bibnamefont
  {Kleinert}}\ and\ \bibinfo {author} {\bibfnamefont {V.}~\bibnamefont
  {Schulte-Frohlinde}},\ }\href@noop {} {\emph {\bibinfo {title} {{Critical
  properties of phi**4-theories}}}}\ (\bibinfo {year} {2001})\BibitemShut
  {NoStop}%
\bibitem [{\citenamefont {Swanson}(2004)}]{swanson2004transition}%
  \BibitemOpen
  \bibfield  {author} {\bibinfo {author} {\bibfnamefont {Mark~S}\ \bibnamefont
  {Swanson}},\ }\bibfield  {title} {\enquote {\bibinfo {title} {Transition
  elements for a non-hermitian quadratic hamiltonian},}\ }\href
  {https://doi.org/10.1063/1.1640796} {\bibfield  {journal} {\bibinfo
  {journal} {Journal of Mathematical Physics}\ }\textbf {\bibinfo {volume}
  {45}},\ \bibinfo {pages} {585--601} (\bibinfo {year} {2004})}\BibitemShut
  {NoStop}%
\bibitem [{\citenamefont {Graefe}\ \emph {et~al.}(2015)\citenamefont {Graefe},
  \citenamefont {Korsch}, \citenamefont {Rush},\ and\ \citenamefont
  {Schubert}}]{1751-8121-48-5-055301}%
  \BibitemOpen
  \bibfield  {author} {\bibinfo {author} {\bibfnamefont {Eva-Maria}\
  \bibnamefont {Graefe}}, \bibinfo {author} {\bibfnamefont {Hans~Jürgen}\
  \bibnamefont {Korsch}}, \bibinfo {author} {\bibfnamefont {Alexander}\
  \bibnamefont {Rush}}, \ and\ \bibinfo {author} {\bibfnamefont {Roman}\
  \bibnamefont {Schubert}},\ }\bibfield  {title} {\enquote {\bibinfo {title}
  {Classical and quantum dynamics in the (non-hermitian) swanson oscillator},}\
  }\href {http://stacks.iop.org/1751-8121/48/i=5/a=055301} {\bibfield
  {journal} {\bibinfo  {journal} {Journal of Physics A: Mathematical and
  Theoretical}\ }\textbf {\bibinfo {volume} {48}},\ \bibinfo {pages} {055301}
  (\bibinfo {year} {2015})}\BibitemShut {NoStop}%
\bibitem [{\citenamefont {Sinha}\ and\ \citenamefont
  {Roy}(2007)}]{1751-8121-40-34-015}%
  \BibitemOpen
  \bibfield  {author} {\bibinfo {author} {\bibfnamefont {A}~\bibnamefont
  {Sinha}}\ and\ \bibinfo {author} {\bibfnamefont {P}~\bibnamefont {Roy}},\
  }\bibfield  {title} {\enquote {\bibinfo {title} {Generalized {S}wanson models
  and their solutions},}\ }\href
  {http://stacks.iop.org/1751-8121/40/i=34/a=015} {\bibfield  {journal}
  {\bibinfo  {journal} {Journal of Physics A: Mathematical and Theoretical}\
  }\textbf {\bibinfo {volume} {40}},\ \bibinfo {pages} {10599} (\bibinfo {year}
  {2007})}\BibitemShut {NoStop}%
\bibitem [{\citenamefont {Aspelmeyer}\ \emph {et~al.}(2014)\citenamefont
  {Aspelmeyer}, \citenamefont {Kippenberg},\ and\ \citenamefont
  {Marquardt}}]{aspelmeyer2014cavity}%
  \BibitemOpen
  \bibfield  {author} {\bibinfo {author} {\bibfnamefont {Markus}\ \bibnamefont
  {Aspelmeyer}}, \bibinfo {author} {\bibfnamefont {Tobias~J}\ \bibnamefont
  {Kippenberg}}, \ and\ \bibinfo {author} {\bibfnamefont {Florian}\
  \bibnamefont {Marquardt}},\ }\bibfield  {title} {\enquote {\bibinfo {title}
  {Cavity optomechanics},}\ }\href {\doibase 10.1103/RevModPhys.86.1391}
  {\bibfield  {journal} {\bibinfo  {journal} {Reviews of Modern Physics}\
  }\textbf {\bibinfo {volume} {86}},\ \bibinfo {pages} {1391} (\bibinfo {year}
  {2014})}\BibitemShut {NoStop}%
\bibitem [{\citenamefont {Gigan}\ \emph {et~al.}(2006)\citenamefont {Gigan},
  \citenamefont {B\"ohm}, \citenamefont {Paternostro}, \citenamefont {Blaser},
  \citenamefont {Langer}, \citenamefont {Hertzberg}, \citenamefont {K.},
  \citenamefont {Baeuerle}, \citenamefont {Aspelmeyer},\ and\ \citenamefont
  {Zeilinger}}]{Gigan2006}%
  \BibitemOpen
  \bibfield  {author} {\bibinfo {author} {\bibfnamefont {S.}~\bibnamefont
  {Gigan}}, \bibinfo {author} {\bibfnamefont {H.~R.}\ \bibnamefont {B\"ohm}},
  \bibinfo {author} {\bibfnamefont {M.}~\bibnamefont {Paternostro}}, \bibinfo
  {author} {\bibfnamefont {F.}~\bibnamefont {Blaser}}, \bibinfo {author}
  {\bibfnamefont {G.}~\bibnamefont {Langer}}, \bibinfo {author} {\bibfnamefont
  {J.~B.}\ \bibnamefont {Hertzberg}}, \bibinfo {author} {\bibfnamefont
  {Schwab}\ \bibnamefont {K.}}, \bibinfo {author} {\bibfnamefont
  {D.}~\bibnamefont {Baeuerle}}, \bibinfo {author} {\bibfnamefont
  {M.}~\bibnamefont {Aspelmeyer}}, \ and\ \bibinfo {author} {\bibfnamefont
  {A}~\bibnamefont {Zeilinger}},\ }\bibfield  {title} {\enquote {\bibinfo
  {title} {Self-cooling of a micro-mirror by radiation pressure},}\ }\href
  {https://doi.org/10.1038/nature05273} {\bibfield  {journal} {\bibinfo
  {journal} {Nature (London)}\ }\textbf {\bibinfo {volume} {444}},\ \bibinfo
  {pages} {67} (\bibinfo {year} {2006})}\BibitemShut {NoStop}%
\bibitem [{\citenamefont {Law}(1995)}]{law1995interaction}%
  \BibitemOpen
  \bibfield  {author} {\bibinfo {author} {\bibfnamefont {CK}~\bibnamefont
  {Law}},\ }\bibfield  {title} {\enquote {\bibinfo {title} {Interaction between
  a moving mirror and radiation pressure: A {H}amiltonian formulation},}\
  }\href {\doibase 10.1103/PhysRevA.51.2537} {\bibfield  {journal} {\bibinfo
  {journal} {Physical Review A}\ }\textbf {\bibinfo {volume} {51}},\ \bibinfo
  {pages} {2537} (\bibinfo {year} {1995})}\BibitemShut {NoStop}%
\bibitem [{\citenamefont {F.}\ and\ \citenamefont
  {Milburn}(2008)}]{WallsMilburn}%
  \BibitemOpen
  \bibfield  {author} {\bibinfo {author} {\bibfnamefont {Walls~D.}\
  \bibnamefont {F.}}\ and\ \bibinfo {author} {\bibfnamefont {G.~J.}\
  \bibnamefont {Milburn}},\ }\href@noop {} {\emph {\bibinfo {title} {{Q}uantum
  {O}ptics}}}\ (\bibinfo  {publisher} {Springer-Verlag},\ \bibinfo {year}
  {2008})\BibitemShut {NoStop}%
\bibitem [{\citenamefont {Chan}\ \emph {et~al.}(2011)\citenamefont {Chan},
  \citenamefont {Alegre}, \citenamefont {Safavi-Naeini}, \citenamefont {Hill},
  \citenamefont {Krause}, \citenamefont {Gr{\"o}blacher}, \citenamefont
  {Aspelmeyer},\ and\ \citenamefont {Painter}}]{chan2011laser}%
  \BibitemOpen
  \bibfield  {author} {\bibinfo {author} {\bibfnamefont {Jasper}\ \bibnamefont
  {Chan}}, \bibinfo {author} {\bibfnamefont {TP~Mayer}\ \bibnamefont {Alegre}},
  \bibinfo {author} {\bibfnamefont {Amir~H}\ \bibnamefont {Safavi-Naeini}},
  \bibinfo {author} {\bibfnamefont {Jeff~T}\ \bibnamefont {Hill}}, \bibinfo
  {author} {\bibfnamefont {Alex}\ \bibnamefont {Krause}}, \bibinfo {author}
  {\bibfnamefont {Simon}\ \bibnamefont {Gr{\"o}blacher}}, \bibinfo {author}
  {\bibfnamefont {Markus}\ \bibnamefont {Aspelmeyer}}, \ and\ \bibinfo {author}
  {\bibfnamefont {Oskar}\ \bibnamefont {Painter}},\ }\bibfield  {title}
  {\enquote {\bibinfo {title} {Laser cooling of a nanomechanical oscillator
  into its quantum ground state},}\ }\href
  {https://doi.org/10.1038/nature10461} {\bibfield  {journal} {\bibinfo
  {journal} {Nature}\ }\textbf {\bibinfo {volume} {478}},\ \bibinfo {pages}
  {89} (\bibinfo {year} {2011})}\BibitemShut {NoStop}%
\bibitem [{\citenamefont {Safavi-Naeini}\ \emph {et~al.}(2012)\citenamefont
  {Safavi-Naeini}, \citenamefont {Chan}, \citenamefont {Hill}, \citenamefont
  {Alegre}, \citenamefont {Krause},\ and\ \citenamefont
  {Painter}}]{safavi2012observation}%
  \BibitemOpen
  \bibfield  {author} {\bibinfo {author} {\bibfnamefont {Amir~H}\ \bibnamefont
  {Safavi-Naeini}}, \bibinfo {author} {\bibfnamefont {Jasper}\ \bibnamefont
  {Chan}}, \bibinfo {author} {\bibfnamefont {Jeff~T}\ \bibnamefont {Hill}},
  \bibinfo {author} {\bibfnamefont {Thiago P~Mayer}\ \bibnamefont {Alegre}},
  \bibinfo {author} {\bibfnamefont {Alex}\ \bibnamefont {Krause}}, \ and\
  \bibinfo {author} {\bibfnamefont {Oskar}\ \bibnamefont {Painter}},\
  }\bibfield  {title} {\enquote {\bibinfo {title} {Observation of quantum
  motion of a nanomechanical resonator},}\ }\href {\doibase
  10.1103/PhysRevLett.108.033602} {\bibfield  {journal} {\bibinfo  {journal}
  {Physical Review Letters}\ }\textbf {\bibinfo {volume} {108}},\ \bibinfo
  {pages} {033602} (\bibinfo {year} {2012})}\BibitemShut {NoStop}%
\bibitem [{\citenamefont {Teufel}\ \emph {et~al.}(2011)\citenamefont {Teufel},
  \citenamefont {Donner}, \citenamefont {Li}, \citenamefont {Harlow},
  \citenamefont {Allman}, \citenamefont {Cicak}, \citenamefont {Sirois},
  \citenamefont {Whittaker}, \citenamefont {Lehnert},\ and\ \citenamefont
  {Simmonds}}]{teufel2011sideband}%
  \BibitemOpen
  \bibfield  {author} {\bibinfo {author} {\bibfnamefont {JD}~\bibnamefont
  {Teufel}}, \bibinfo {author} {\bibfnamefont {Tobias}\ \bibnamefont {Donner}},
  \bibinfo {author} {\bibfnamefont {Dale}\ \bibnamefont {Li}}, \bibinfo
  {author} {\bibfnamefont {JW}~\bibnamefont {Harlow}}, \bibinfo {author}
  {\bibfnamefont {MS}~\bibnamefont {Allman}}, \bibinfo {author} {\bibfnamefont
  {Katarina}\ \bibnamefont {Cicak}}, \bibinfo {author} {\bibfnamefont
  {AJ}~\bibnamefont {Sirois}}, \bibinfo {author} {\bibfnamefont {Jed~D}\
  \bibnamefont {Whittaker}}, \bibinfo {author} {\bibfnamefont {KW}~\bibnamefont
  {Lehnert}}, \ and\ \bibinfo {author} {\bibfnamefont {Raymond~W}\ \bibnamefont
  {Simmonds}},\ }\bibfield  {title} {\enquote {\bibinfo {title} {Sideband
  cooling of micromechanical motion to the quantum ground state},}\ }\href
  {https://doi.org/10.1038/nature10261} {\bibfield  {journal} {\bibinfo
  {journal} {Nature}\ }\textbf {\bibinfo {volume} {475}},\ \bibinfo {pages}
  {359} (\bibinfo {year} {2011})}\BibitemShut {NoStop}%
\bibitem [{\citenamefont {Bollini}\ and\ \citenamefont
  {Giambiagi}(1986)}]{bollini1986lagrangian}%
  \BibitemOpen
  \bibfield  {author} {\bibinfo {author} {\bibfnamefont {CG}~\bibnamefont
  {Bollini}}\ and\ \bibinfo {author} {\bibfnamefont {JJ}~\bibnamefont
  {Giambiagi}},\ }\href@noop {} {\emph {\bibinfo {title} {Lagrangian procedures
  for higher order field equations}}},\ \bibinfo {type} {Tech. Rep.}\ (\bibinfo
   {institution} {Centro Brasileiro de Pesquisas Fisicas},\ \bibinfo {year}
  {1986})\BibitemShut {NoStop}%
\bibitem [{\citenamefont {Nieto}(1997)}]{NIETO1997135}%
  \BibitemOpen
  \bibfield  {author} {\bibinfo {author} {\bibfnamefont {Michael~Martin}\
  \bibnamefont {Nieto}},\ }\bibfield  {title} {\enquote {\bibinfo {title}
  {Displaced and squeezed number states},}\ }\href {\doibase
  https://doi.org/10.1016/S0375-9601(97)00183-7} {\bibfield  {journal}
  {\bibinfo  {journal} {Physics Letters A}\ }\textbf {\bibinfo {volume}
  {229}},\ \bibinfo {pages} {135 -- 143} (\bibinfo {year} {1997})}\BibitemShut
  {NoStop}%
\end{thebibliography}%
\appendix
\titlespacing*{\section}
{0pt}{18pt}{5pt}
\section{Non-local Hamiltonian from the field theoretic Stress-Energy tensor}\label{SET}
In this appendix we derive the Hamiltonian for the non-local Schr\"odinger field (to first order in the non-locality scale) in three different ways: from the relativistic Lagrangian with subsequent non-relativistic limit, from the non-relativistic Lagrangian via Noether's theorem, and by Legendre transforming the non-relativistic Lagrangian.  

\subsection{Relativistic field as a starting point}
Consider a relativistic complex scalar field described by 
\begin{equation}
    \mathcal{L}=-\frac{1}{2}\left[\phi^*\left(\mathcal{KG}+a_2 \mathcal{KG}^2\right)\phi+\phi\left(\mathcal{KG}+a_2 \mathcal{KG}^2\right)\phi^*\right],
\end{equation}
where $\mathcal{KG}=\Box+m^2$. This Lagrangian can be seen as the first order approximation of \eqref{lag} in the non-locality scale. While doable, we do not need to consider higher order corrections. Following~\cite{bollini1986lagrangian} we can obtain the $T^{00}$ of the theory. The variation of the Lagrangian, $\mathcal{L}\left[\phi,\phi^*,\mathcal{KG}\phi,\mathcal{KG}\phi^*,\mathcal{KG}^2\phi,\mathcal{KG}^2\phi^*\right]$, is given by
\begin{equation}\label{bollini2}
\delta\mathcal{L}=\frac{\partial{\mathcal{L}}}{\partial\phi}\delta\phi+\frac{\partial{\mathcal{L}}}{\partial\mathcal{KG}\phi}\delta\mathcal{KG}\phi+\frac{\partial{\mathcal{L}}}{\partial\mathcal{KG}^2\phi}\delta\mathcal{KG}^2\phi+\dots+c.c.
\end{equation}
The computation is straightforward and leads to
\begin{align}\label{bollini3}
\delta\mathcal{L}=&-\frac{1}{2}\partial_{\mu}\left[\phi^*\delta\partial^{\mu}\phi-\partial^{\mu}\phi^*\delta\phi+a_2 \phi^*\delta\partial^{\mu}\Box\phi\right.\\ \nonumber
&\left.-a_2 \partial^{\mu}\phi^*\delta\Box\phi+a_2 \Box\phi^*\delta\partial^{\mu}\phi-a_2 \partial^{\mu}\Box\phi^*\delta\phi\right.\\ \nonumber
&\left.+2a_2 m^2 \phi^*\delta\partial^{\mu}\phi-2a_2 m^2 \partial^{\mu}\phi^*\delta\phi+ c.c.\right]
\end{align}
From this expression it is a trivial task to obtain the $T^{00}$ for the reletivistic field. However, taking the non-relativistic limit is less trivial. The reason being that a naive non-relativistic limit of the Hamiltonian density leads to unphysical divergences. The root of this problem lies in the way in which the variation has been performed. In particular, after having separated the rest-energy phase in the field as $\phi=\psi e^{-imc^2 t}$, we need to require that the variation symbol in \eqref{bollini3} does not act on the rest-energy phase. The final result is a Hamiltonian denisity given by
\begin{equation}\label{a4}
    \mathcal{H}=-\frac{1}{2m}\psi^*\left(\nabla^2-a_2 \nabla^4\right)\psi-\frac{1}{2m}\psi\left(\nabla^2+a_2 \nabla^4\right)\psi^*-4a_2 m\dot{\psi^*}\dot{\psi}
\end{equation}

\subsection{The non-relativistic Lagrangian as a starting point}
An equivalent way to arrive at the same result is to start from the non-relativistic limit of the non-local Lagrangian:
\begin{equation}
    \mathcal{L}=\psi^* \mathcal{S}\psi-2 a_{2} m \psi^* \mathcal{S}^2 \psi +\psi \mathcal{S}^*\psi^*-2 a_{2} m \psi \mathcal{S}^{*2} \psi^*.
\end{equation}
We follow again the general recipe of ref.\cite{bollini1986lagrangian}, and look at the variation of the Lagrangian for infinitesimal translations in order to compute the stress-energy tensor of the higher order theory. Let us write the Lagrangian as
\begin{equation}\label{dep}
    \mathcal{L}=\mathcal{L}(\psi,\psi^*,\mathcal{S}^{n}\psi,\mathcal{S}^{*n}\psi^*),
\end{equation}
then we can write the variation of the Lagrangian as
\begin{equation}\label{bollini}
\delta\mathcal{L}=\frac{\partial{\mathcal{L}}}{\partial\psi}\delta\psi+\frac{\partial{\mathcal{L}}}{\partial\mathcal{S}\psi}\delta\mathcal{S}\psi+\frac{\partial{\mathcal{L}}}{\partial\mathcal{S}^2\psi}\delta\mathcal{S}^2\psi+\dots+c.c.,
\end{equation}
and compare it with the effect of a translation on a scalar
\begin{align}
    \delta\mathcal{L}&=\left(\mathcal{S}-2 a_{2} m \mathcal{S}^2\right) \psi \delta\psi^*+\left(\mathcal{S}-2 a_{2} m \mathcal{S}^*2\right) \psi^* \delta\psi\\ \nonumber
    &+\psi^*\delta\mathcal{S}\psi+\psi\delta\mathcal{S}^*\psi^*-2 a_{2} m \psi \delta\mathcal{S}^{*2} \psi^*-2 a_{2} m \psi^* \delta\mathcal{S}^2 \psi\\ \nonumber
    &= \partial_{t}\left(i\psi^*\epsilon^{\mu}\partial_{\mu}\psi-i\psi\epsilon^{\mu}\partial_{\mu}\psi^*\right)+\\ \nonumber
    & -\frac{1}{2}\partial_{x}\left(\partial_{x}\psi^*\epsilon^{\mu}\partial_{\mu}\psi+\partial_{x}\psi\epsilon^{\mu}\partial_{\mu}\psi^*-\psi^*\epsilon^{\mu}\partial_{x}\partial_{\mu}\psi-\psi\epsilon^{\mu}\partial_{x}\partial_{\mu}\psi^*\right)\\ \nonumber
    & +S^*\psi^* \delta\psi+S\psi \delta\psi^*-2a_{2}m\psi^*\delta\left(\frac{1}{4m^2}\nabla^4 \psi+\frac{i}{m}\nabla^{2}\dot{\psi}-\ddot{\psi}\right)\\ \nonumber
    &-2a_{2}m\psi\delta\left(\frac{1}{4m^2}\nabla^4 \psi^*-\frac{i}{m}\nabla^{2}\dot{\psi^*}-\ddot{\psi^*}\right)\\ \nonumber
    &=\epsilon^{\mu}\partial_{\mu}\mathcal{L}.
\end{align}
Integrating by parts and using the equations of motion we obtain the Hamiltonian density
\begin{equation}\label{a9}
    \mathcal{H}=-\frac{1}{2m}\psi^*\left(\nabla^2-a_2 \nabla^4\right)\psi-\frac{1}{2m}\psi\left(\nabla^2+a_2 \nabla^4\right)\psi^*-4a_2 m\dot{\psi^*}\dot{\psi},
\end{equation}
in accordance with \eqref{a4}.

Finally, adding a potential $V(x)$ gives
\begin{align}\label{hamil}
    H=&\int d^3 x\left[-\frac{1}{2m}\psi^*\left(\nabla^2-a_2 \nabla^4\right)\psi-\frac{1}{2m}\psi\left(\nabla^2-a_2 \nabla^4\right)\psi^*\right.\\ \nonumber
    &\left.-4a_2 m\dot{\psi^*}\dot{\psi}+V(x)\psi^*\psi\right].
\end{align}

\subsection{Legendre transform and Hamilton's equations of motion}
Consider again the Lagrangian in the non-relativistic limit
\begin{equation}
    \mathcal{L}=\psi^* \left(S-2a_2 mS^2\right)\psi+\psi \left(S^*-2a_2 mS^{*2}\right)\psi^*.
\end{equation}
We can easily reduce this Lagrangian to first order in time  by integrating by parts. (It should be noted that, in what follows, the reduction to first order in time is not necessary. Indeed, using the Ostrogradski method for higher derivatives theories, it is possible to define canonical coordinates and the Hamiltonian starting from the second order in time Lagrangian.) This is possible since a total derivative in the Lagrangian does not influence the dynamics of the system. The reduced Lagrangian is given by
\begin{align}
    \mathcal{L}=&i\psi^* \dot{\psi}+\frac{1}{2m}\psi^*\nabla^2 \psi-\frac{a_2}{2m}\psi^* \nabla^{4}\psi-2a_2 i\nabla^2\psi^* \dot{\psi}-2a_2 m\dot{\psi^*}\dot{\psi}\\ \nonumber
    &-i\psi \dot{\psi^*}+\frac{1}{2m}\psi\nabla^2 \psi^*-\frac{a_2}{2m}\psi \nabla^{4}\psi^*+2a_2 i\nabla^2\psi \dot{\psi^*}-2a_2 m\dot{\psi^*}\dot{\psi}
\end{align}
We define canonical coordinates and canonical momenta
\begin{align}
& Q=\psi\\ \nonumber
& P=\frac{\delta\mathcal{L}}{\delta\dot{\psi}}=i\psi^*-2a_2 i\nabla^2\psi^* -4a_2 m\dot{\psi}^*\\ \nonumber
\end{align}
together with their complex conjugates.

The Hamiltonian is obtained by Legendre transforming the Lagrangian:
\begin{align}\label{hqqdot}
    \mathcal{H}&=P\dot{Q}+P^*\dot{Q^*}-\mathcal{L}\\ \nonumber
    &= -\frac{1}{2m}\psi^*\nabla^2 \psi-\frac{1}{2m}\psi \nabla^2 \psi^*+\frac{a_2}{2m}\psi^*\nabla^4 \psi+\frac{a_2}{2m}\psi\nabla^4 \psi^*-4a_2 m\dot{\psi}^*\dot{\psi},
\end{align}
in accordance with \eqref{a4} and \eqref{a9}. 

It can be useful to express the Hamiltonian solely in terms of canonical coordinates and momenta. This can be done solving for the canonical momenta with respect to $\dot{\psi}^*,\dot{\psi}$, and rewriting the Lagrangian as
\begin{align}
   \mathcal{L}=&P \dot{\psi}+P^*\dot{\psi}^*+4a_2\dot{\psi}^* \dot{\psi}\\ \nonumber
   &+\frac{1}{2m}\psi^*\nabla^2 \psi-\frac{a_2}{2m}\psi^* \nabla^{4}\psi+\frac{1}{2m}\psi\nabla^2 \psi^*-\frac{a_2}{2m}\psi \nabla^{4}\psi^* . 
\end{align}
The Hamiltonian is then given by
\begin{align}\label{hpq}
    \mathcal{H}(Q,P)=&-\frac{1}{2m}\psi^*\nabla^2 \psi+\frac{a_2}{2m}\psi^* \nabla^{4}\psi-\frac{1}{2m}\psi\nabla^2 \psi^*\\ \nonumber
    &+\frac{a_2}{2m}\psi \nabla^{4}\psi^*-\left(\frac{\psi^*\psi}{4a_2 m}-\frac{1}{2m}\psi\nabla^2\psi^*-\frac{1}{2m}\psi^*\nabla^2\psi\right.\\ \nonumber
    &\left.+\frac{i}{4a_2 m}P\psi-\frac{i}{4a_2 m}P^*\psi^*+\frac{a_2}{m}\nabla^2\psi\nabla^2\psi^*\right.\\ \nonumber
    &\left.-\frac{i}{2m}\nabla^2\psi P+\frac{i}{2m}P^*\nabla^2\psi^*+\frac{P^*P}{4a_2 m}\right),
\end{align}
and Hamilton's equations are
\begin{align}
    & \dot{P}=-\frac{\delta\mathcal{H}}{\delta Q}=-\left[\frac{\partial\mathcal{H}}{\partial\psi}+\nabla^2\frac{\partial\mathcal{H}}{\partial\nabla^2\psi}+\nabla^4\frac{\partial\mathcal{H}}{\partial\nabla^4\psi}\right]\\ \nonumber
    & \dot{Q}=\left[\frac{\delta\mathcal{H}}{\delta P}\right],
\end{align}
and their complex conjugates. Note that the first Hamilton equation gives (twice) the second order Schr\"odinger equation for $\psi^*$, as expected. The second Hamilton equation instead gives $\dot{\psi}=\dot{\psi}$. In an analogous way, the complex conjugates equations lead to the non-local Schr\"odinger equation for $\psi$ and another tautology.

\section{Time-dependent perturbation theory}\label{TimeDepPert}
In this Appendix we report the details of the calculations with time-dependent perturbation theory.

\subsection{Displaced Perturbed Ground State}
The perturbed ground state $|\Omega\rangle$ has been obtained in~\eqref{pstate} from time-independent perturbation theory. Let us rewrite it in terms of the energy eigenstates of the harmonic oscillator $|n\rangle$
\begin{equation}
    |\Omega\rangle=|0\rangle+\epsilon (a|2\rangle+b|4\rangle),
\end{equation}
where the coefficients $a,\,b$ are given in~\eqref{pstate}. We displace this state by acting with the displacement operator $\mathcal{D}[\alpha]$,
\begin{equation}\label{disp}
    \mathcal{D}[\alpha]|\Omega\rangle=|\alpha\rangle+\epsilon\left(a|2,\alpha\rangle+b|4,\alpha\rangle\right).
\end{equation}
Here $|n,\alpha\rangle$ are the displaced number states (see e.g.,~\cite{NIETO1997135}) which can be represented as
\begin{equation}
    |n,\alpha\rangle=e^{-|\alpha|^2/2}\sum_{k=0}^{\infty}\sum_{j=0}^{n}C^{n}_{k,j}|n+k-j\rangle,
\end{equation}
where 
\begin{equation}
    C^{n}_{j,k}=\frac{\alpha^k}{k!}\frac{(-\alpha*)^j}{j!}\left[\frac{(n-j+k)!n!}{(n-j)!(n-j)!}\right]^{1/2}.
\end{equation}
We can rewrite these states more compactly as
\begin{equation}
    |n,\alpha\rangle=e^{-|\alpha|^2/2}\sum_{k}b^{(n)}_{k}|k\rangle,
\end{equation}
where the coefficients $b^{(n)}_k$ for $n=2,4$ are given by
\begin{align}
    & b^{(2)}_{0}=C^2_{2,0}\\
    & b^{(2)}_{1}=C^2_{1,0}+C^2_{2,1}\\
    & b^{(2)}_{2}=C^2_{0,0}+C^2_{1,1}+C^2_{2,2}\\
    & b^{(2)}_{n>2}=C^2_{0,n-2}+C^2_{1,n-1}+C^2_{2,n},\\
    & b^{(4)}_{0}=C^4_{4,0}\\
    & b^{(4)}_{1}=C^4_{3,0}+C^4_{4,1}\\
    & b^{(4)}_{2}=C^4_{2,0}+C^4_{3,1}+C^4_{4,2}\\
    & b^{(4)}_{3}=C^4_{1,0}+C^4_{2,1}+C^4_{3,2}+C^4_{4,3}\\
    & b^{(4)}_{4}=C^4_{0,0}+C^4_{1,1}+C^4_{2,2}+C^4_{3,3}+C^4_{4,4}\\
    & b^{(4)}_{n>4}=C^4_{0,n-4}+C^4_{1,n-3}+C^4_{2,n-2}+C^4_{3,n-1}+C^4_{4,n}.
\end{align}
We can rewrite~\eqref{disp} in the energy eigenstates basis of the unperturbed harmonic oscillator in terms of these coefficients
\begin{align}\label{disstate}
    \mathcal{D}[\alpha]|\Omega\rangle &=e^{-|\alpha|^2/2}\sum_{n=0}^{\infty}\left(\frac{\alpha^n}{\sqrt{n!}}+\epsilon a b_{n}^{(2)}+\epsilon b b_{n}^{(4)}\right)|n\rangle\\
    &=\sum_{n=0}^{\infty}b_{n}|n\rangle
\end{align}

Now consider~\eqref{perteq} and assume~\eqref{disstate} is the initial state. We see that $a^{(0)}_n(0)$ are nothing but the coefficients of the standard coherent state $|\alpha\rangle$, while $a^{(1)}_n(0)=\exp{(-|\alpha|^2/2)}(a\,b_{n}^{(2)}+b\,b_{n}^{(4)})$. Using 
the matrix elements in~\eqref{matrixe} we can write
\begin{align}\label{statet}
    |\psi(t)\rangle&=e^{-|\alpha|^2/2}\sum_{n=0}^\infty \left(\frac{\alpha^n}{\sqrt{n!}}+\epsilon(a\,b^{(2)}_{n}+b\,b^{(4)}_{n})-i\epsilon d_n(t)\right)e^{-i E_n t}|n\rangle\\
    &=\sum_{n=0}^\infty a_n(t)e^{-i E_n t}|n\rangle,
\end{align}
where 
\begin{align}
    d_{n}(t)&=\sum_{k=0}^{\infty}a^{(0)}_k(0)\int_0^t dt' \langle n|\mathcal{H}|k\rangle e^{i(n-k)t'}
\end{align}

\subsection{Mean value and Variance of a general quadrature}
Having the form of the state in the energy basis of the harmonic oscillator we can now compute the mean and variance for several observables.
Before doing so, let us note that the time-dependent coefficients $d_{n}(t)$ contain in general secular terms linear in time. The same feature was already discussed in~\cite{PhysRevLett.116.161303,PhysRevD.95.026012}. In the following these terms are suppressed.  

Consider the general quadrature $x(\theta)=\frac{1}{\sqrt{2}}(\hat{a}e^{-i\theta}+\hat{a}^{\dag}e^{i\theta})$. The mean value and mean squared value over the state~\eqref{statet} are given by 
\begin{align}
    &\langle x(\theta)\rangle = \sum_{n=0}^\infty \left(e^{i(\theta+t)}a^*_{n+1} a_n+e^{-i(\theta+t)}a^*_n a_{n+1}\right)\sqrt{\frac{n+1}{2}},\label{meanvar1}\\ 
    &\langle x(\theta)^2\rangle = \sum_{n=0}^\infty \left(e^{2i(\theta+t)}a^*_{n+2} a_n+e^{-2i(\theta+t)}a^*_n a_{n+2}\right)\frac{1}{2}\sqrt{\frac{(n+1)!}{n!}}\label{meanvar2}\\ 
    &+\sum_{n=0}^\infty |a_n|^2\left(\frac{1}{2}+n\right),\nonumber
\end{align}
where we have used the fact that 
\begin{align}
    &\langle n|x(\theta)|m\rangle =\frac{1}{\sqrt{2}}\left(e^{i\theta}\sqrt{1+m}\delta_{1+m,n}+e^{-i\theta}\sqrt{m}\delta_{1-m,-n}\right),
\end{align}
and
\begin{align}
    \langle n|x(\theta)^2|m\rangle = \frac{1}{2}& \left(e^{-2i\theta}\sqrt{m}\sqrt{m-1}\delta_{m,n+2}\right.\\ \nonumber
   &\left.+e^{2i\theta}\sqrt{m+2}\sqrt{m+1}\delta_{2+m,n}\right)+\left(\frac{1}{2}+m\right)\delta_{n,m}.
\end{align}
The variance is then simply $\langle x(\theta)\rangle^2-\langle x(\theta)^2\rangle$.
\\
\\
\indent It should be noted that the above expressions also includes the cases in which the initial state is the perturbed ground state, a standard coherent state, and the standard vacuum state. Indeed, setting $\alpha=0$ we recover the case in which the initial state is the perturbed ground state $|\Omega\rangle$. Additionally setting $a=b=0$, we recover the harmonic oscillator's ground state $|0\rangle$ as initial state. And, finally, setting $\alpha\neq 0$ and $a=b=0$ we obtain as initial state the coherent state $|\alpha\rangle$. Thus~\eqref{meanvar1},\eqref{meanvar2} are general enough to derive all the results discussed in the main text.

\end{document}